\documentclass[tighten,twocolumn]{aastex62}
\usepackage[]{graphicx}
\usepackage{amsmath}
\usepackage[normalem]{ulem}

\shortauthors{KERR}
\shorttitle{Variability Analyses with Photon Weights}

\begin{document}

\title{Multi-scale time- and frequency-domain likelihood analysis with photon
weights}

\author[0000-0002-0893-4073]{M.~Kerr}
\affiliation{Space Science Division, Naval Research Laboratory, Washington, DC 20375--5352, USA}
\correspondingauthor{M.~Kerr}
\email{matthew.kerr@nrl.navy.mil}

\begin{abstract}
We present an unbinned likelihood analysis formalism employing photon
weights---the probabilities that events are associated with a
particular source.  This approach is applicable to any
photon-resolving instrument, and thus well suited to high-energy
observations; we focus here on GeV $\gamma$-ray data from the
\textit{Fermi} Large Area Telescope.  Weights connect individual
photons to the outputs of a detailed, expensive likelihood analysis of
a much larger data set.  The weighted events can be aggregated into
arbitrary time spans ranging from microseconds to years.  Such
retrospective grouping permits time- and frequency-domain analysis
over a wide range of scales and enables characterization of disparate
phenomena like blazar flares, $\gamma$-ray bursts, pulsar pulses,
novae, $\gamma$-ray binaries, and other variable sources.  To
demonstrate the formalism, we incorporate photon weights into the
Bayesian blocks algorithm and perform a hierarchical time scale
analysis of 3C~279 activity.  We analyze pulsar pulse profiles and
estimate the unpulsed emission level and the optimal division of the
data into on- and off-pulse intervals.  We extend the formalism to
Fourier analysis and derive estimators for power spectra, used to
search for and characterize periodic sources.  We show how the Fast
Fourier transform can be used to probe orbital periods as short as a
minute and we discuss mitigation of spurious signals.  Our final
example combines time- and frequency- domain analysis to jointly
characterize the flares and orbital modulation of Cygnus X-3, yielding
the strongest detection of the orbital signal ($>$13$\sigma$) to date.
Finally, we discuss extensions of the work to other GeV sources and to
X-ray and TeV observations.
\vspace{1cm}
\end{abstract}


\section{Introduction}

The detection and analysis of transients and variable sources, in both
the time and frequency domains, is increasingly important to advancing
the frontiers of astrophysics.  Time-domain astronomy with non-imaging
instruments---including most $\gamma$-ray telescopes, some X-ray
telescopes, and radio and gravitational wave interferometers---is
challenging because these instruments do not directly measure source
intensities.  Instead, models of backgrounds and sources in the field
are folded through the instrument response and compared to the
observed data, typically using maximum likelihood and/or Bayesian
inference to determine model parameters.  Such analysis often requires
an expensive optimization or exploration of a high-dimension parameter
space.  Since by definition a new source is absent from existing
models, searches for faint or transient sources demand an iterative
approach where candidate sources are trialled over a range of possible
parameters.

The situation is complicated further by source confusion, either from
overlapping point sources or from a strong, diffuse background.  MeV
and GeV $\gamma$-ray telescopes fall into this r\'{e}gime, as the
reconstruction of the direction of incident $\gamma$ rays through
Compton scattering or $e^-$/$e^+$ pair production produces typical angular
resolutions from 0.1--10\arcdeg.  Furthermore, the MeV and GeV sky glows
brightly with the diffuse emission of cosmic rays
interacting with
the Galactic matter and radiation fields, and it boasts thousands of
point sources, primarily blazars and pulsars \citep[e.g.][]{4FGL}.
This unique challenge makes the analysis of $\gamma$-ray data
especially suitable for the methods we develop below, and accordingly
we concentrate on data from the \textit{Fermi} Large Area Telescope
\citep[LAT,][]{Atwood09}.

Because the LAT has few-degree angular resolution at 100\,MeV,
self-consistent analysis of a single LAT source requires modelling 
all sources within and near to a relatively large region of interest
(ROI; say 10$\arcdeg$)  with good precision.  This is a computationally
intensive procedure, and modelling the diffuse background is a complex
task.  When performing the analysis, it is necessary to choose the data
boundaries in time, energy, and space at the outset, and typically
source models describe the time-averaged intensity only.  To maximize
sensitivity, the entire LAT data range (now over 10 years) may be
co-added.  This approach is followed by the \textit{Fermi}-LAT team in
producing point source catalogs and leads to the most detailed model
of the $\gamma$-ray sky.  Conversely, to characterize
rapid variability of the brightest sources, short spans of data might
be binned into 6-, 12-, or 24-hour intervals.  In other words, the
choice of binning determines the measurements that can be made, and
blinds the observer to variability on time scales that aren't well
suited to the binning.

Some approaches aimed at overcoming this problem have appeared in the
literature.  \citet{Lott12} describe a method for adaptively defining
``constant error'' bins within each of which flux of a target source
can be determined with similar precision.  \citet{Scargle98} outlines
a more general algorithm, now widely known as ``Bayesian blocks'', in
which data are divided into fundamental \textit{cells} which are built
up into \textit{blocks} in a way that optimally partition the data
according to some fitness function, e.g. the posterior Poisson
likelihood.  All of the cells within a block are supposed to have the
same source intensity/rate.  The smallest possible cells are the
intervals between photon arrivals, so this approach can probe a wide
range of time scales.  The algorithm faces some challenges, however,
with LAT data.  With photon-based cells and a Poisson likelihood
fitness function, it may lose sensitivity if the background is strong.
(On the other hand, this works very well during bright gamma-ray
bursts.) Using longer cells, e.g.  monthly data segments, can increase
sensitivity by allowing more sophisticated fitness functions, like the
multi-dimension likelihood discussed above.  However, information on
shorter time scales is lost.

In this work, we describe a method of \textit{retrospective} analysis
in which the information from a full likelihood analysis, performed
over the entire \textit{Fermi}-LAT data span for maximum sensitivity,
is encoded into photon \textit{weights}, the probability for each
photon to have originated from a particular source.  Photon weights
have been suggested \citep{Bickel08} and adopted \citep{Kerr11} as a
means to enhance sensitivity to $\gamma$-ray pulsations.  More
recently, \citet{Bruel19} described a method for computing weights for
faint sources, but we note those weights do not
incorporate the same information as the probability-based formulation
adopted here.

The fundamental idea is that each weight captures information about
the ratio of source-to-background rates at a given time and energy,
and that changes in source properties \textit{relative to the mean}
can then be inferred by comparing the distribution of weights within
subsets of the data to that of the whole. If a blazar flares one day,
we will observe relatively more photons with high weights on that day.
If its spectrum hardens, we will observe relatively more weights at
high energy.  On the other hand, if the distribution of weights within
subsets of the data is indistinguishable from the mean distribution,
the source is not variable.  We put this notion on a formal footing
below.  Like the Bayesian blocks algorithm, the method is limited only
by photon counting statistics, but it explicitly incorporates
background information even at the photon level.  Because much of the
work is done in the initial likelihood analysis, the retrospective
analysis outlined below imposes little additional computational
burden.  It is thus well suited for exploratory analysis and the
identification of periods of interest which can then be

We conclude this introductory discussion by noting that the methods
developed here are distinct from ``weighted likelihood'' analysis
\citep[e.g.][]{HZ2002}, in which the contributions of various subsets
of the data to the total likelihood are replaced by a weighted sum:
$\log\mathcal{L}=\sum_i\log\mathcal{L}_i\rightarrow\log\mathcal{L}_w\equiv\sum_iw_i\log\mathcal{L}_i$.
Small weights can be used to decrease the influence of data that may
be affected by systematic errors or otherwise fail to follow the
probability density function assumed by $\mathcal{L}$.  And indeed,
the \textit{Fermi}-LAT team used weighted likelihood in the 4FGL
analysis to moderate the effect of uncertainties in the diffuse
background model \citep{4FGL}.  The weights in this work are not ad
hoc but derived from a perfectly standard, albeit expensive,
likelihood analysis, and they encapsulate information used in
approximate but standard, fast, and flexible likelihood applications.
More technically, the weights here are not applied directly to
contributions to the log likelihood, but rather appear ``inside'' of
the log.

In the next section, we derive the formalism, and in the sequel we
focus on two applications.  First, in \S\ref{sec:lc} we demonstrate
the estimation of light curves of both blazars and pulsars (viz. pulse
profiles) and we show how the Bayesian blocks algorithm using maximum
likelihood can be applied to optimally identify and characterize
multi-time scale variability.  Second, in \S\ref{sec:ps}, we show 
how to use the formalism for Fourier analysis.  We derive the
``exposure weighted'' power spectra of \citet{Corbet07}, used to
search LAT data for $\gamma$-ray binaries, including the normalization
for trials factor correction.  We also show how to compute these
estimators with the Fast Fourier transform, enabling searches for very
short-period binaries.  In \S\ref{sec:reweighting}, we combine both
time-domain and frequency-domain techniques to simultaneously
characterize the slow flares (weeks) and fast orbital modulation
(hours) of Cygnus X-3, producing the strongest detection of this
modulation to date.  Finally, we conclude in \S\ref{sec:discussion}
with a summary and suggestions for other applications of the methods.

\onecolumngrid
\section{General Formalism}
\label{sec:formalism}
A general likelihood $\mathcal{L}$ for a typical high-energy instrument, in
which uncorrelated photons from a variety of sources are dispersed in both position ($\Omega$) and energy ($E$), is
\begin{displaymath}
\log\mathcal{L} = \sum_i \bigg[ n_i \log \sum_j \lambda_j(\Omega_i,
E_i) -\sum_j \lambda_j(\Omega_i, E_i) \bigg],
\end{displaymath}
where the outer $i$-sum is over position and energy bins 
and the inner $j$-sum is over all the sources considered in the ROI.
The expected counts $\lambda_j$ arise from folding a model for the
$j$th source through the instrument response to predict the counts in
bin $i$, including the effects of varying exposure.
If these bins are taken to be so small that $n_i\in(0,1)$, this
becomes the ``unbinned'' likelihood \citep{Tompkins99},
\begin{align*}
\log\mathcal{L} =& \sum_i \bigg[ \log \sum_j \lambda_j(\Omega_i,
E_i) -\sum_j  \lambda_j(\Omega_i,E_i)\bigg]\\
\equiv& \sum_i \bigg[ \log \sum_j \lambda_{j,i}\bigg] -\sum_j \Lambda_j,
\end{align*}
\noindent
with $\Lambda$ giving the total rate summed over all the ROI.  Let us
assume that we have optimized the model parameters such that
$\mathcal{L}$ is maximized and the $\lambda_j$ give the time-averaged
rates for each source.  Now, suppose we
partition the data into arbitrary segments and consider the $k$th
segment $P_k$.  If the exposure varies, we define the exposure factor
$f_k$ as the fraction of the total exposure in $P_k$, and
to encapsulate source variability let us define
$\lambda_{j,k}=(1+\alpha_{j,k})\lambda_j$.  Then, the log likelihood
for $P_k$, the $k$th segment is
\begin{align*}
\log\mathcal{L}_k =& \sum_{i\in P_k} \bigg[ \log \sum_j f_k
(1+\alpha_{j,k})\lambda_{j,i}\bigg]- f_k \sum_j(1+\alpha_{j,k}) \Lambda_j\\
=&\sum_{i\in P_k} \bigg[ \log \sum_j (1+\alpha_{j,k})\lambda_{j,i}\bigg]
- f_k \sum_j\alpha_{j,k} \Lambda_j +\sum_{i\in P_k} \log f_k - f_k \sum_j
\Lambda_j\\
=&\sum_{i\in P_k} \bigg[ \log \sum_j (1+\alpha_{j,k})\lambda_{j,i}\bigg]
- f_k \sum_j\alpha_{j,k} \Lambda_j + const.
\end{align*}
For simplicity, we have suppressed dependence of $\alpha$
on energy, but the method presented here can also account for spectral
evolution, and we refer the reader to \citet{Guillemot19} for examples
with phase-resolved pulsar spectroscopy.
Now, let us introduce the notion of a photon \textit{weight}, the
probability a particular photon originated from a particular source.
It is simply the predicted rate relative to the total rate,
$w_{j,i}\equiv\lambda_{j,i}/\sum_j\lambda_{j,i}$.  In terms of these
weights, we can write the likelihood as
\begin{align}
\nonumber\log\mathcal{L}_k =&\sum_{i\in P_k} \bigg[ \log \bigg(1
+\sum_j \alpha_{j,k}
\frac{\lambda_{j,i}}{\sum_m\lambda_{m,i}}\bigg)\bigg]
- f_k \sum_j\alpha_{j,k} \Lambda_j +\sum_{i\in P_k} \bigg[ \log \sum_m
\lambda_{m,i}\bigg]\\
\label{eq:general_like}
\equiv&\sum_{i\in P_k} \bigg[ \log\bigg( 1 + \sum_j
\alpha_{j,k}w_{j,i}\bigg)\bigg]
- f_k \sum_j\alpha_{j,k} \Lambda_j + const.
\end{align}
This formulation is the heart of the methods presented here.  The
$\alpha_{j,k}$ describe how source intensities vary over time and the
goal of this work is to efficiently and reliably estimate these
quantities by (re) optimizing the likelihood in Eq.
\ref{eq:general_like}.  Because
the photon weights are calculated once and for all using the results
of the global likelihood analysis, we no longer need to worry about
the instrument properties and can optimize $\log\mathcal{L}_k$
directly---a substantial reduction in complexity.  The remainder of
this work is the development and application of methods of estimating
$\alpha$.

To streamline this development, we now impose a major simplifying
assumption.  First, we focus on a particular source, say
$\alpha\equiv\alpha_0$.  Next, we suppose that the variations of other
sources in the ROI can be modelled with a single background term, i.e.
$\sum_{j>0} \alpha_{j} w_{j,i}\equiv \beta \sum_{j>0} w_{j,i}=
\beta\,(1-w_{0,i})$.  With this assumption, the log likelihood becomes
\begin{align}
\nonumber
\log\mathcal{L}_k =&\sum_{i\in P_k} \bigg[ \log \bigg(1 + \alpha_{0,k}
w_{0,i} + \sum_{j>0} \alpha_{j,k}w_{j,i}\bigg)\bigg]
- f_k \alpha_{0,k}\Lambda_0 - f_k \sum_{j>0}\alpha_{j,k} \Lambda_j\\
\nonumber
\approx&
\sum_{i\in P_k} \bigg[ \log \bigg(1 + \alpha_{0,k}
w + \beta_k \sum_{j>0} w_{j,i}\bigg)\bigg]
- f_k \alpha_{0,k} \Lambda_{0,k} - f_k \beta_k \sum_{j>0} \Lambda_{j,k}\\
\label{eq:simple_like}
\equiv&
\sum_{i\in P_k} \bigg[ \log \bigg(1 + \alpha_k
w_i + \beta_k(1-w_i)\bigg)\bigg]
- \alpha_k S_k - \beta_k B_k.
\end{align}
As an additional simplification, we note that we can often replace the
total predicted model counts, $S_k$ and $B_k$, with the estimators
$S_k=f_k\sum_{i}\,w_i$ and $B_k=f_k\sum_{i}\,(1-w_i)$, i.e. the total
weighted sums scaled by the exposure factors.  These are good
estimators so long as there are several hundred photons in the total
data set, almost always the case for \textit{Fermi}-LAT analysis.

\newpage
\twocolumngrid
With this simplified form, our task is reduced to determining two
quantities, the source and background amplitudes, $\alpha_k$ and
$\beta_k$, for each interval $P_k$.  Although we expect sources to
vary independently (e.g.  background blazars) or not at all (e.g.
diffuse Galactic emission), in practice we find this simplifying
assumption is good because: (1) when confusion is important (e.g. at
low energies), the spatial distribution of weights is broad and varies
weakly over the ROI (2) if a background source varies strongly enough
to affect the ROI, it may also dominate the background contribution;
(3) the use of a small ROI minimizes the effects of more distant
sources.  Although the remainder of this work will use Eq.
\ref{eq:simple_like}, we emphasize that an analyst can instead adopt
Eq. \ref{eq:general_like} for more complicated cases when it is
advisable to model multiple background sources.

Following a brief discussion of data preparation and software tools,
we proceed to develop two applications of Eq. \ref{eq:simple_like}: the
characterization of light curves by estimating piecewise-constant
$\alpha_k$ and $\beta_k$, and the characterization of Fourier
amplitudes by analyzing sinusoidal variation of $\alpha$ and $\beta$.

\begin{figure}
\centering
\vspace{0.2cm}
\includegraphics[angle=0,width=0.98\linewidth]{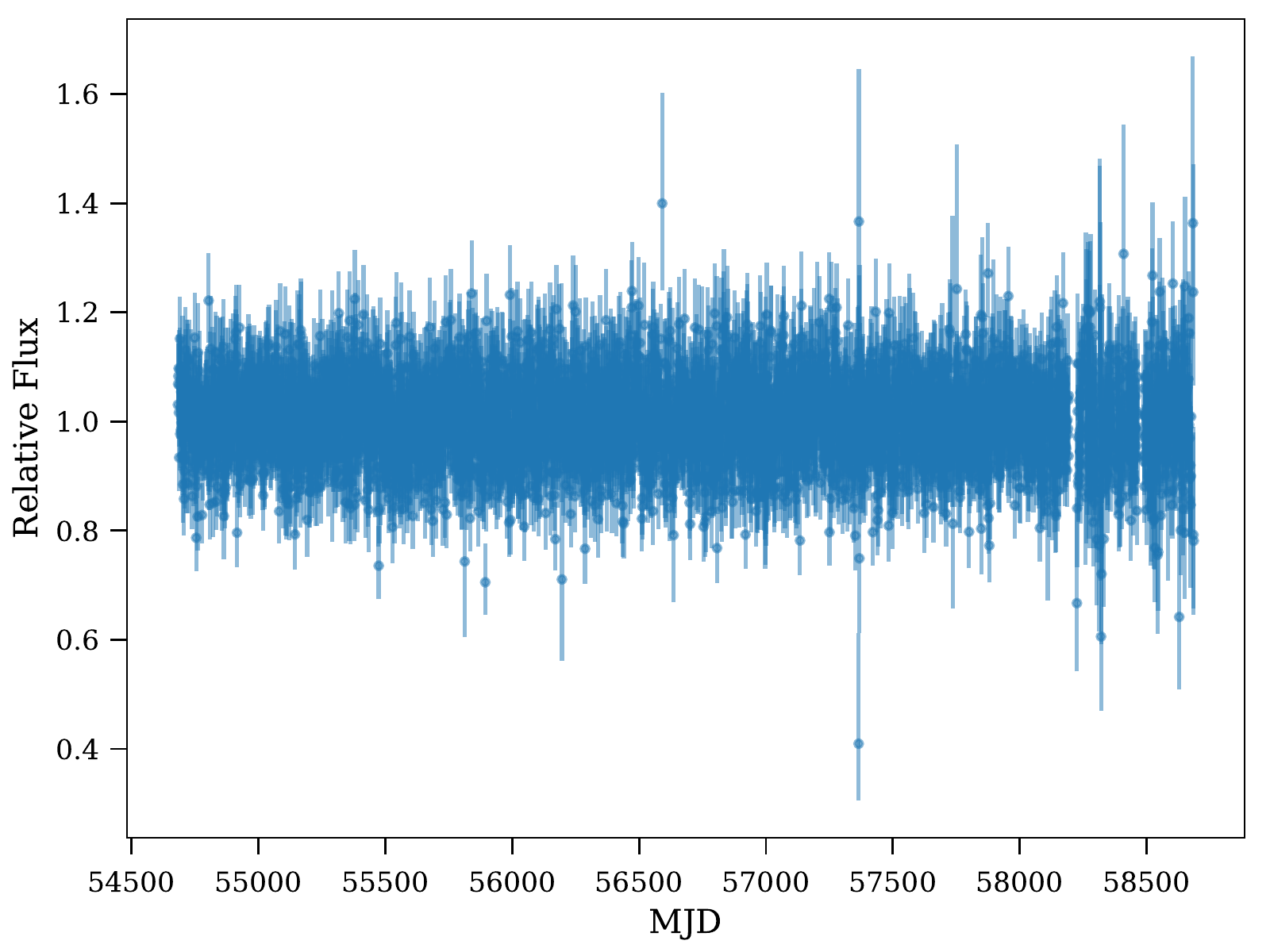}
\caption{\label{fig:fig1_geminga}Light curve for Geminga  with
1-day resolution. Days with exposure below threshold are not depicted, and the source is otherwise strongly detected on each interval.}
\end{figure}

\begin{figure*}
\centering
\includegraphics[angle=0,width=0.98\textwidth]{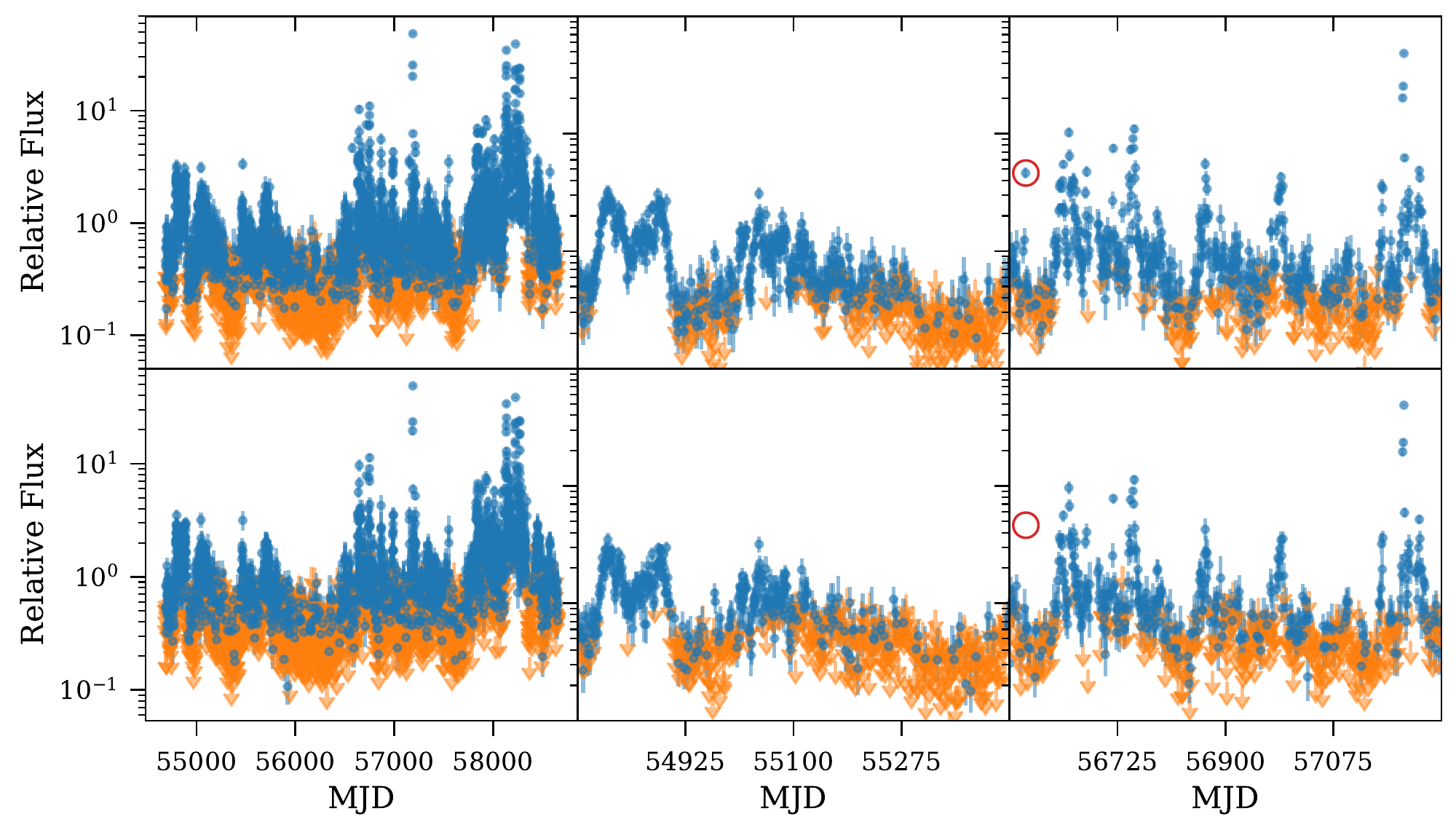}
\caption{\label{fig:fig1_3c279}Light curve for 3C~279.  The upper
panels show estimates made with the background fixed to its average
value ($\beta=0$), while the lower panels are computed with a profile
likelihood maximized with respect to $\beta$.  Upper limits (orange
arrows) are shown for days with TS $<9$.  The middle and right panels
focus on 700-day intervals beginning MJD~54750 and 56550,
respectively.  The measurement indicated by the red circle corresponds
to an M1.5 solar flare (see main text), which is confused with the
blazar.  The profile likelihood correctly assigns this flux to the
background.}
\end{figure*}

\section{Weight Computation, Exposure Calculation, and Software}

For all examples presented here, we use about 10 years of Pass 8
\citep{Atwood13} \textit{Fermi}-LAT data, selected to have
reconstructed arrival directions within 2$\arcdeg$ (typically),
5$\arcdeg$, or 10$\arcdeg$ (as noted) of the source.  We compute
photon weights using the \textit{pointlike} \citep{Kerr10} application
and a model of the sky based on the
FL8Y\footnote{\url{https://fermi.gsfc.nasa.gov/ssc/data/access/lat/fl8y/}}
source list.


To compute the expected source rates in a given time interval, it is
necessary to know the exposure---the product of effective area and
integration time---towards the source.  In a typical likelihood
analysis, e.g. one carried out with the \textit{Fermi} Science Tool
\textit{gtlike}, the exposure is
calculated as a function of energy and position and spectral analyses
are carried out over large (say 20$\arcdeg$) ROI.  Because we
are using the products of such an analysis and restricting attention
to a small ROI, we are able we evaluate the exposure only at the
position of the source.  We further use a spectral model (by default a
power law with spectral index $-2.1$) to average the effective area
over energy.  In principle, better results are obtained by using the
correct spectral model for the source, but in practice, we find little
difference compared to this omnibus model.
We compute the exposure directly from the 30-s intervals
tabulated in the FT2 file distributed by the Fermi Science Support
Center\footnote{\url{https://fermi.gsfc.nasa.gov/ssc/data/access/}}.

We have made the software to compute exposure and to perform all of
the analyses presented here available in the package
\texttt{godot}\footnote{\url{https://github.com/kerrm/godot}}.  In
particular, \texttt{godot} provides routines for the selection of LAT
data on a variety of time scales, e.g. contiguous viewing periods or
uniform time bins, and aggregating the data and exposure into
arbitrarily-sized cells as required for further analysis.  Some care
has been taken to optimize the performance.  Routines for computing
data cells using Solar System barycentric time, critical for
identifying very short period binaries, are also provided (see
\S\ref{sec:ps}).

Therefore, in summary, users wishing to try out the techniques
described here should (1) install the \textit{Fermi} Science Tools and
\texttt{godot}; (2) adapt an existing sky model (e.g. 4FGL) or develop
a new one using $\textit{gtlike}$ if the source of interest is not in the sky model (3) use
the sky model and the \textit{gtsrcprob} Science Tool to produce photon weights for all
sources of interest; (4) use the applications in \texttt{godot} to
apply retrospective likelihood analysis.

\section{Light Curves}
\label{sec:lc}

A \textit{light curve} is a time series of point estimators of the
intensity of emission received from a given source.  From Eq.
\ref{eq:simple_like}, we see we can obtain such estimators by dividing the data into
suitable segments $P_k$ and maximizing
$\mathcal{L}_k(\alpha_k,\beta_k)$.  We can estimate
confidence intervals by applying a uniform prior (restricted to
positive values, $\alpha_k,\beta_k$$>$$-1$) and computing the range which encloses a specified
fraction of the posterior
distribution for $\alpha_k$/$\beta_k$ (68\% for 1\,$\sigma$ error flags) .  In determining $P_k$, for practicality, we restict
attention to segments of $\geq$30\,s, which is the interval at which
the spacecraft (S/C) position and orientation are recorded in standard data
products. Even the brightest sources, with
fluxes (E\,$>$\,100\,MeV) of $10^{-5}$ to $10^{-4}$
ph\,cm$^{-2}$\,s$^{-1}$, are barely detectable in such short segments,
and a more useful interval is the roughly 20 minutes during which the
source is in the LAT field-of-view during a typical (zenith-pointed)
orbit.  Besides gamma-ray bursts, only a few sources have been
observed to show variability at or below the orbital time-scale, e.g.
flaring blazars like 3C~279 \citep{Ackermann16_3C279}, 3C~454.3
\citep{Abdo11_3C454.3}, and 4C$+$21.43 \citep{Tanaka11}---see also
\citet{Meyer19} for a more general analysis; the third
observed periastron passage of PSR~B1259$-$63 \citep{Johnson18}; and
the Crab Nebula \citep[e.g.][]{Buehler12,Abdo11_crab}.  Accordingly,
analyses typically use longer (1 day, 1 week) bins.  But by selecting
a longer time-scale for averaging, there is risk of washing out rapid
variability.  As we will show below, adopting relatively short bins
for the likelihood analysis and using aggregating techniques like
Bayesian blocks or multi-scale filtering is an effective means of
probing a wide range of time scales.

\subsection{Examples and Profile Likelihood}

Here, we consider two sources as examples.  First, we analyze the
Geminga pulsar (PSR~J0633$+$1746), the second-brightest pulsar in the
\textit{Fermi} sky \citep{Abdo10_geminga}.  To date, it is not known
to show any variability, and it thus makes a good test source for
probing systematic errors in the method.  To estimate its light curve,
we take $\alpha_k$ as constant over a single day, and we fix
$\beta=0$.  To avoid measurements with large uncertainty, we exclude
days whose exposure is $<$10\% of the mean.
The resulting estimators (maximum
likelihood with 1\,$\sigma$ error bars) appear in Figure
\ref{fig:fig1_geminga}.  As expected, they are consistent with
$\alpha_k=0$ to a high degree of accuracy.  A failure in one of the
\textit{Fermi} solar panel array drives (MJD~58194) resulted in a
revised S/C rocking profile and less homogeneous exposure on
few-week timescales.  The resulting inhomogeneity in measurement
precision following this date is apparent.  Nonetheless, the
uncertainties are accurately estimated and the error-normalized
measurements follow an outlier-free unit normal distribution.

As a second example, we consider the highly variable blazar 3C~279 and
as above we compute daily estimators of the flux density.  Unlike
Geminga, 3C~279 is not always bright enough to detect every day, and
we report upper limits when the TS$\equiv2\times[
\log\mathcal{L}(\hat{\alpha_k})-\log\mathcal{L}(-1)]$ (with
$\hat{\alpha_k}$ the maximum likelihood estimator) is $<9$.  The light
curve, displayed in Figure \ref{fig:fig1_3c279}, shows relatively
typical (albeit intense) flares lasting multiple days, in keeping with
causality arguments about source size.  The leftmost panels shows the
entire data range, while the right panels focus on shorter, 700-day
intervals to show more structure.  In the right panel, one
measurement, indicated with a red circle at MJD 56576.5, seems to be
an ``orphan'' flare surrounded by upper limits.  Such variability is
faster than expected, and indeed this point is actually due to a
behind-the-limb solar flare producing bright MeV--GeV emission, a fast
coronal mass ejection, and solar energetic particles
\citep{Ackermann17_solar_flares}.  At the time of this flare, the sun
was only 2$\arcdeg$ from 3C~279, and their emission is confused,
leading to the incorrect flux estimator.  In the bottom panels, we
show the same light curves obtained by maximizing the profile
likelihood, $\log\mathcal{L}(\alpha_k,\hat{\beta_k}(\alpha_k))$, the
likelihood maximized with respect to the background for each value of
$\alpha_k$.  Because the extended solar emission over the ROI
``looks'' very much like an additional, flat background component, the
new degree of freedom absorbs the contribution from the solar flare
and the best-fit flux value and TS drop from $F/F_{\mathrm{mean}}=4.6$
and TS$=342$ to $F/F_{\mathrm{mean}}$=0 and TS$=0$, i.e. a
non-detection.  In general, the profile likelihood requires modestly
more computation and reduces measurement precision.  Here, the typical
TS is halved, though this depends on ROI size (see Fig.
\ref{fig:ls5039_aperture}).  But it clearly provides an important check
on the $\beta=0$ assumption, and future work could improve measurement
precision by adopting a more physical prior on $\beta$.

\subsection{Bayesian Blocks}

\begin{figure}
\centering
\includegraphics[angle=0,width=0.98\linewidth]{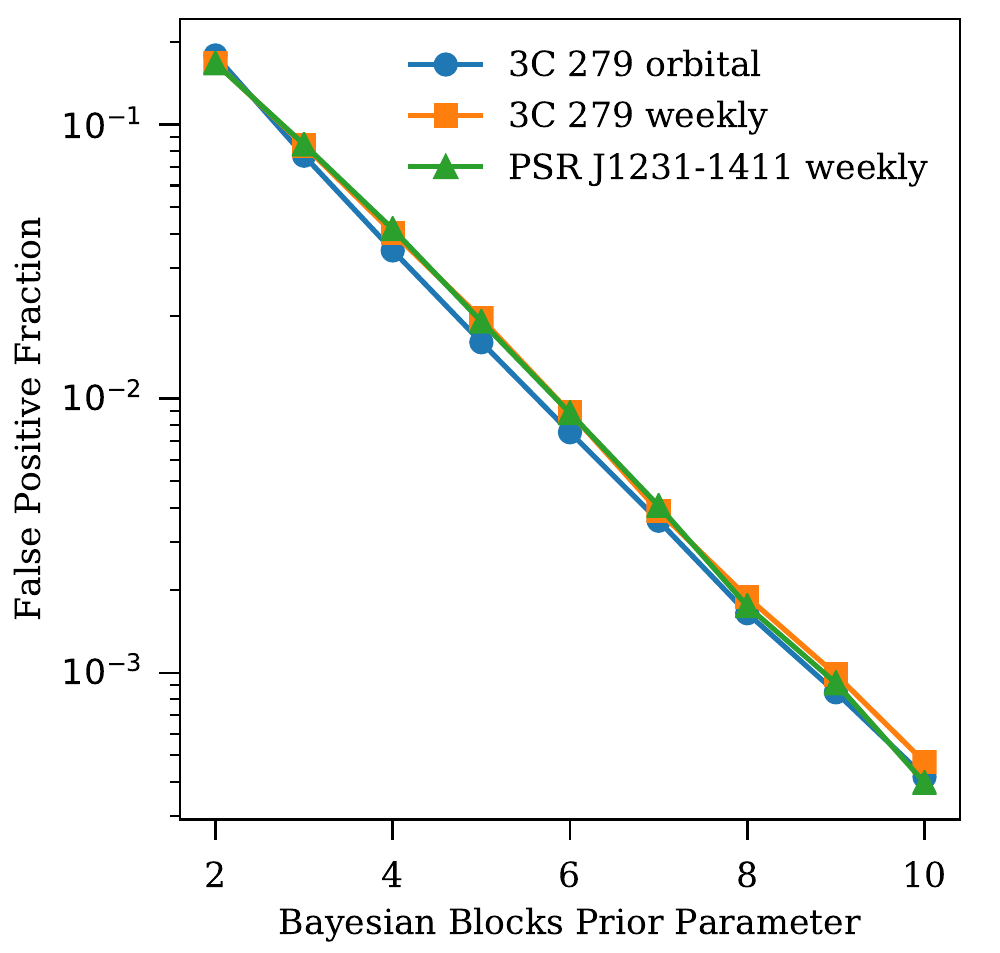}
\caption{\label{fig:fig5}Simulations showing the false positive rate
(defined as the number of blocks in excess of one divided by the total
number of cells) as a function of the prior parameter $\gamma$.}
\end{figure}

Bayesian blocks \citep[hereafter ``BB'',][]{Scargle98} is a method of
partitioning a data set comprising $N_c$ elementary cells into $N_b$
longer blocks that maximize the posterior probability according to
some model of variability.  Most simply, this model is piecewise
constant, and the method can be viewed as a way of compressing cells
into blocks of nearly-constant emission level.  The level of
compression is controlled by the priors adopted for the variability
model, and in the piecewise constant model it can be taken as a prior
on $N_b$.  A commonly-used prior, which we adopt, takes the form
$\pi(N_b)\propto N_b^{-\gamma}$, penalizing additional blocks with
``strength'' controlled by $\gamma$.

With this prior, or any prior such the log posterior for independent
data segments is additive, there is an efficient $\mathcal{O}(N_c^2)$
dynamic programming algorithm for determining the optimal data
partition \citep{Scargle13}.  This approach is particularly attractive
when paired with a fast likelihood algorithm, in which case it enables
multi-scale variability analysis by providing the algorithm with many
very short data cells and allowing it to ``detect'' variability via a
preference for one or more change points.  If the optimal partition
has only one block, there is no variability.  In the presence of
variability, the algorithm automatically and optimally divides the
data up such that the measurement precision is high while fast
variability is not oversmoothed by binning.

The BB algorithm starts with a single cell (which is already the
optimal partition!) and with each iteration adds a new cell and
identifies a new optimal partition.  Because there will be $N_c$
iterations, each iteration must be completed in
$\mathcal{O}(N_c)$ operations to achieve the
$\mathcal{O}(N_c^2)$ complexity.  Generally, the $j$th iteration will require evaluation of $j$ new
fitness functions on partitions containing on average $j/2$ cells and
$\mathcal{O}(j)$ photons.  If the fitness function is independent of
the partition length, e.g. the Poisson likelihood for total counts,
the $j$ $\mathcal{O}(1)$ operations satisfy the $\mathcal{O}(j)$
complexity for the iteration.  But evaluating Eq. \ref{eq:simple_like}
for a typical cell requires $\mathcal{O}(j)$ operations, giving
$\mathcal{O}(j^2)$ complexity for the iteration and
$\mathcal{O}(N_c^3)$ overall.  To avoid this, we have implemented
a caching scheme in which we first analyze the log likelihood for each
cell to determine the location of its maximum ($\hat{\alpha}$) and the
points at which it has decreased by a given amount (by default 30).
We then evaluate and store the likelihood on a grid of points over
this range (by default $N_{pt}=200$), and we can optimize the
likelihood for the sequence of blocks simply by combining these grids
and finding the maximum.  In this way, evaluation of the fitness
function for the sequence requires $\mathcal{O}(N_c\times
N_{pt})$ operations.  This scheme is implemented in \texttt{godot} and
used for the the following analyses.

Sensitivity to variability, e.g. the false positive and false negative
rate, is governed by the prior.  Our choice of $N_b^{-\gamma}$ reduces
the log likelihood by $\gamma$ for each block (degree of freedom),
enforcing parsimony.  Generally, $\gamma$ must be carefully chosen to
match the properties of the specific data set.  When the individual
data cells are sufficiently large that the Central Limit Theorem
applies, the log likelihoods will follow $\chi^2$ distributions.
Then, the false positive rate (creation of a spurious change point)
will depend only on $\gamma$ and $N_c$ and not on the properties of
the source in question.  Then $\gamma$ can be trivially tuned to give
the desired false positive rate.  On the other hand, if some cells
have only a few photons of low weight, as might be the case for those
comprising single \textit{Fermi}-LAT orbits, the false positive rate
would depend sensitively on the exact exposure.  In this case,
simulations are useful in determining the ideal prior.

\begin{figure}
\centering
\includegraphics[angle=0,width=0.95\linewidth]{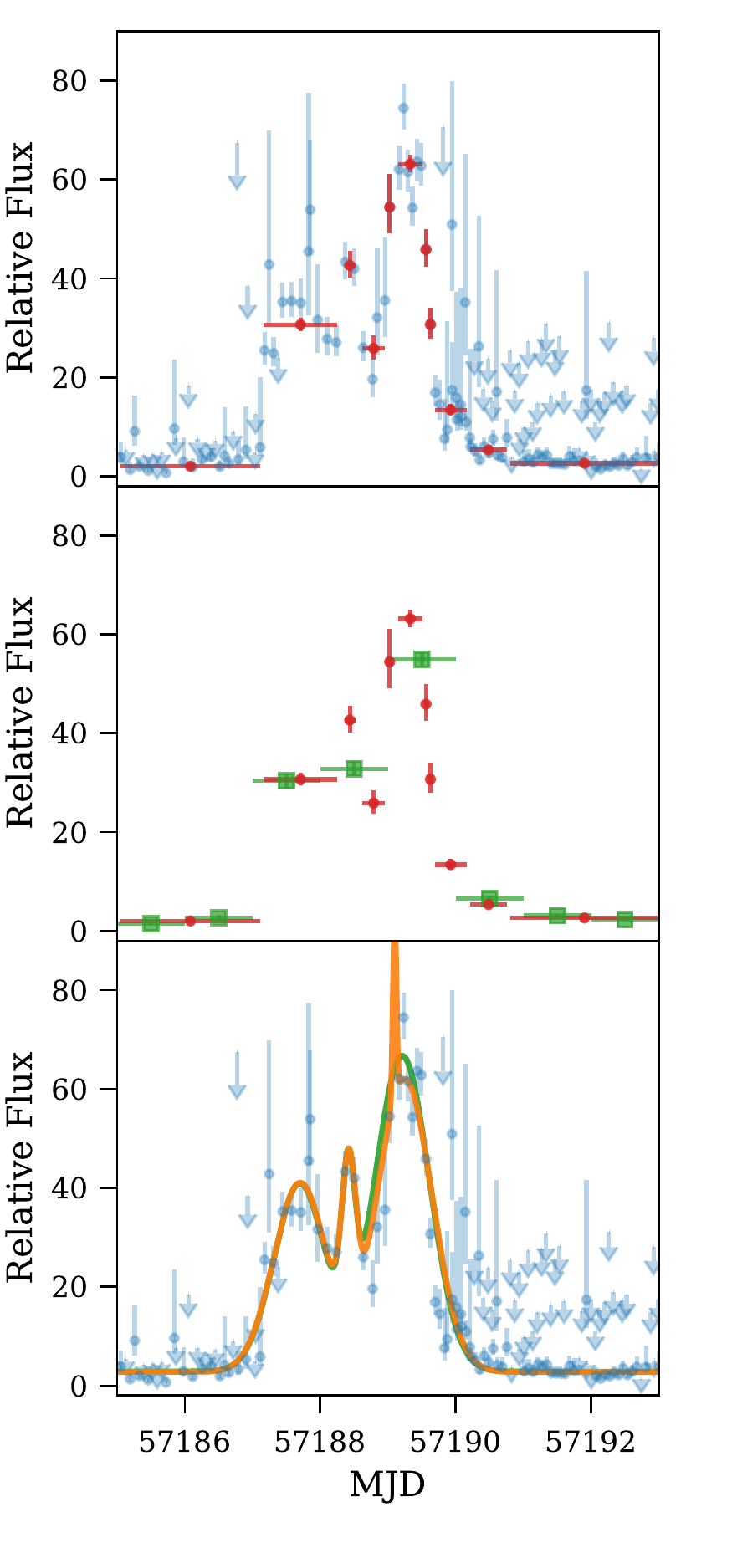}
\caption{\label{fig:3c279_flare}  Light curve for a single 3C~279 flare.
(Top) The raw likelihood fits from individual orbits are shown in
blue; the typical duration of the data acquired during an orbit is too
narrow to illustrate.  Blocks obtained with the BB algorithm are shown
in red, with horizontal errors indicating the block extent and
vertical errors the flux uncertainty.  
(Middle) BB shown in red as in top.  The green points indicate
likelihood fits performed with 1\,d integrations, with flux density
errors too small to see.
(Bottom) As top, blue points show likelihood fits from individual
orbits, while two analytic models with 3 (green) and 4 (orange)
gaussian are fit using the orbital likelihoods.}

\end{figure}

We have carried out such simulations by re-distributing photons
randomly amongst cells according to their computed source rates and
then applying the BB algorithm.  (We have included this simulation
capability in \texttt{godot}.)  Some of our test cases are illustrated
in Figure \ref{fig:fig5} and include uniform bins and steady sources
(PSR~J1231$-$1411, one-week bins), uniform bins with strongly variable
sources (3C~279, one-week bins), and nonuniform, very short bins with
strongly variable sources (3C~279, orbital/$\sim$20-minute bins).
Interestingly, we find that the false positive rates for all test
cases closely follow the exponential shape of the prior, regardless of
fluctuation level of the cells.  The rate is roughly
$0.8\,\exp(-0.8\gamma)$, reminiscent of the aymptotic null
distribution of the $H$ statistic \citep{Kerr11}, which similarly
involves a maximization over $\chi^2$ variables.  This calibration
allows a sensible choice of $\gamma$ for any data set.  Many typical
analyses are covered by a dynamic range of $\sim$1000, so selecting
$\gamma=8$ with a false positive rate $\sim$0.001 works well.  Wider
dynamic ranges (e.g. searches for sub-day variability in the full
Fermi data set) require $\gamma \geq10$.

To summarize this discussion, \textbf{the combination of a fast
likelihood and Bayesian blocks is a powerful, universal method for
detecting variability on a wide range of time scales}.  We demonstrate
this with an analysis of a bright flare from 3C~279, using data from
MJD 57185 to 57193.  As cells, we select photons from contiguous
exposure intervals, typically about 20\,m every 3\,hr (two orbits),
though exposure interruptions from passage through the South Atlantic
Anomaly can produce two shorter exposures.  In this case, a Target of
Opportunity request resulted in several days of pointed mode
observations, so in total we have 146 cells over the 8\,d of data.
Accordingly we choose $\gamma=8$ for the prior.  The top panel of
Figure \ref{fig:3c279_flare} shows the results for both the flux
density estimates from individual cells (blue) and the resulting BB
partition (red).  In its brightest state, 3C~279 is strongly detected in single
orbits, while observations outside of the flare yield only upper
limits.  The BB algorithm accurately identifies all of the major
variability evident in the high time resolution data, including a
sharp sub-flare between the two broad peaks and the very strong peak
lasting one single orbit.  Moreover, by aggregating the data in the
off-peak intervals, it provides good estimates of the quiescent flux.
By contrast, the middle panel shows the same data set but binned into
1\,d cells (green), a common choice for monitoring blazars.  The main
peak is under resolved, the rapid variability between and within the
peaks is smoothed away, and quiescent portions are broken up over
multiple bins.

\begin{figure*}
\centering
\includegraphics[angle=0,width=0.49\linewidth]{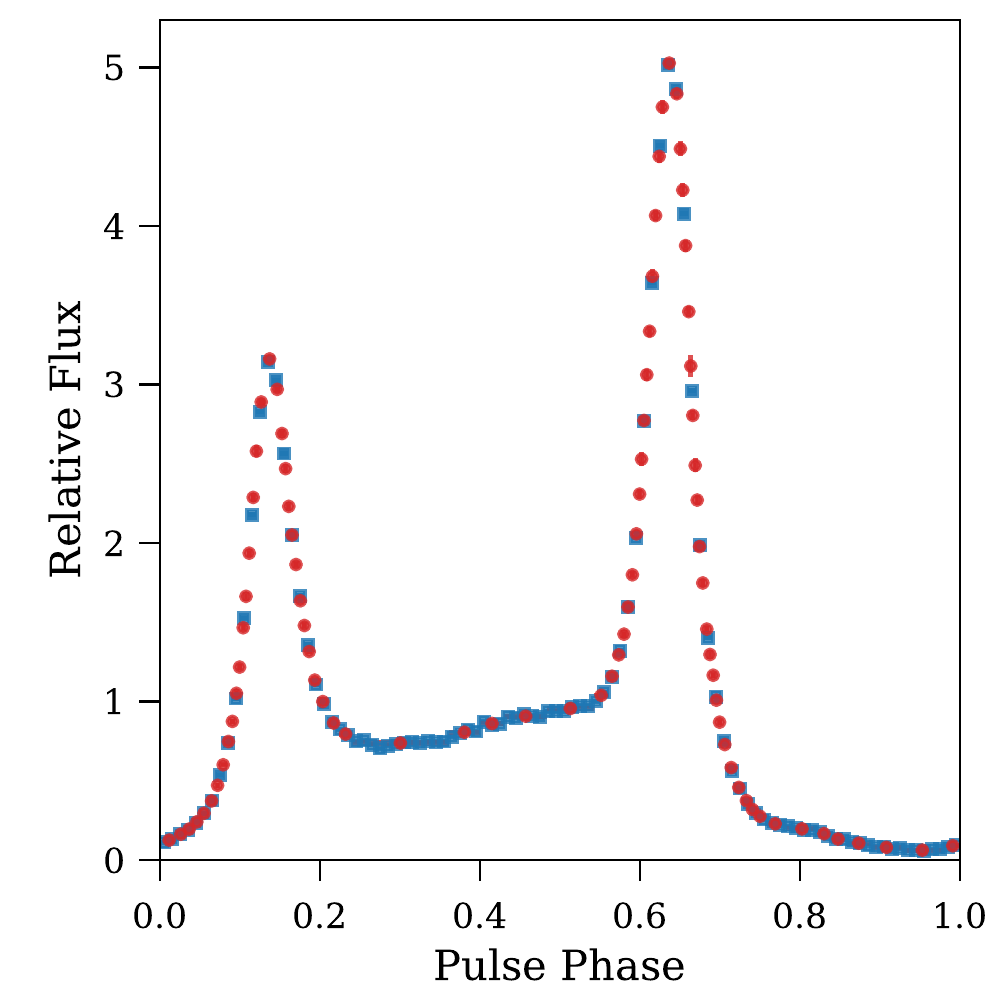}
%
\includegraphics[angle=0,width=0.49\linewidth]{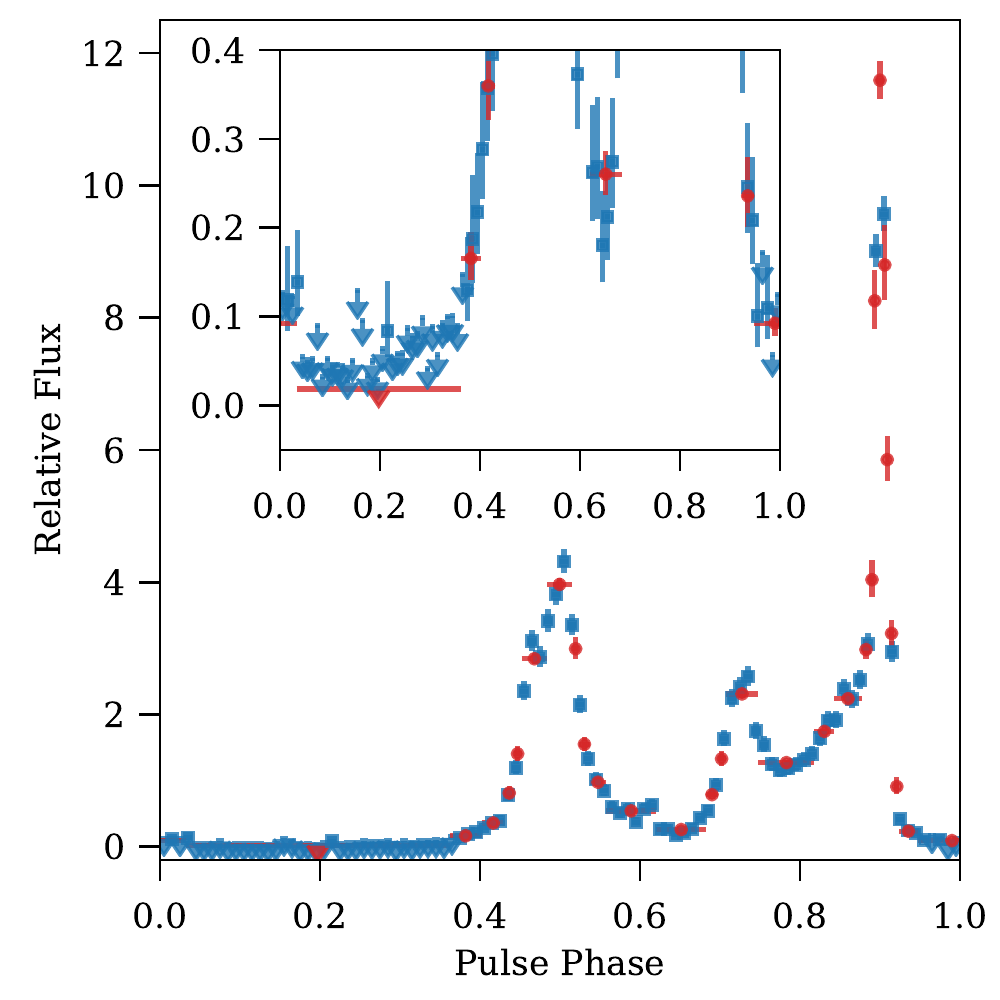}
\caption{\label{fig:fig4}Pulse profiles for Geminga (left) and
PSR~J1231$-$1411 (right).  Blue points give maximum likelihood estimates in 100
uniform phase bins, while red symbols indicate the BB partition.  The inset emphasizes the stringent off-peak upper limit.}
\end{figure*}

\subsection{Waveforms}
\label{sec:waveforms}

Although we have concentrated on piecewise models, we include here an
example of fitting an analytic shape using the orbital-time scale
likelihood.  At these time scales, the flux density estimators are
highly non-gaussian, so it is better to use the full likelihood rather
than, e.g., using least squares to fit a functional form to the
estimators.
Applications might include characterizing rise and fall
times of flares, investigating the significance of transient features,
and model selection.  Here, we investigate the presence of fast flares
by first fitting, using maximum likelihood, a simple three-gaussian
model to the apparent three-peak structure of the flare.  The
resulting model appears in green in the bottom panel of Figure
\ref{fig:3c279_flare}, along with the direct orbital measurements.
These orbital measurements appear to indicate an un-modelled fast
flare at the onset of the third peak.  To test the significance of
this feature, we add and optimize a fourth gaussian, which improves
the log likelihood by 20, indicating it is likely to be a real
feature.  Although this example is entirely ad hoc, we direct the
reader to \citet{Meyer19} for physically motivated modelling of fast
variability in 3C~279 and other blazars.


\subsection{Pulse Profiles}

The flux density of pulsars as a function of rotational phase (also
widely referred to as ``light curves'', but here called pulse profiles
for clarity) are particularly well suited to this
method.  Indeed, because the rotational time scale is so well separated
from typical times over which exposure and backgrounds vary,
any background variation is common across pulse phase bins
and does not affect pulse profile estimation.  Pulse profiles are
typically presented as histograms, either of Stokes parameters
recorded by radio telescopes or of photons by high-energy experiments.
Consequently, we adopt the piecewise-constant formulation for the source intensity.  However, we note that unbinned analytic models,
e.g. truncated Fourier expansions, gaussians, lorentzians, etc. can
also be used to model $\alpha$ (see \S\ref{sec:waveforms}).

The likelihood formalism here has several advantages over previous
approaches.  The most suitable direct comparison are the weighted
photon histograms presented in \citet{Abdo13_2pc}, which have
estimators for the background level and bin uncertainty; these
quantities are both estimated from the global weights distribution,
and certain artifacts appear where the bin distribution differs
substantially, e.g. a narrow off-pulse or a bright peak.  In contrast, the likelihood method here yields a
direct estimate both of the pulsar flux density and of its uncertainty
based entirely on the ``local'' distribution of photons.  It is, in
essence, a fixed-shape but \textit{bona fide} spectral analysis of
each bin.  (See \citet{Guillemot19} for a more complete treatment of
phase-resolved spectroscopy.)

Moreover, this approach admits a fitness function appropriate for the
BB algorithm.  Previous analyses applying BB to pulse profiles have
either used only unweighted counts \citep[e.g.][]{Lande14}, losing
sensitivity, have used less powerful statistics like the $F$-test
as a fitness function \citep{Caliandro13}, or require a transformation
of the raw data to a more suitable form \citep{Ajello19}.  As in the previous section, we can use an
intrinsically high resolution, say 1000 phase bins, for computation of
the data partition, while keeping a modest resolution, say 100 phase
bins, for a descriptive presentation.  As before, we use simulations
to determine the correct BB prior to give $\sim$1
partition in the case of uniform phase data with the same distribution
of weights.  (We remind the reader here that, in order to enforce the
periodic boundary condition of a pulse profile, the BB
algorithm should be applied to three full rotations, with the data
representation taken from the partition of the central rotation.)

For our first example, we again use the Geminga pulsar, and the
now-familiar scheme appears in the left panel of Figure
\ref{fig:fig4}.  The blue points result from a ``typical'' pulse
profile of 100 uniform phase bins, while the red points are a
BB partition based on 1000 bins.  The analyis shows the ad hoc binning
has underresolved the two peaks, but also clearly demonstrates that
there is no fine structure in the pulse profile.  As a second example,
we consider the bright millisecond pulsar J1231$-$1411
\citep{Ransom11}, which has one of the sharpest peaks of Fermi
pulsars.  The 100-bin profile dramatically underresolves the main
peak, while at the same time overresolves the off-pulse.  By correctly
aggregating the bins with the BB method, we capture both the
milliperiod structure in the bright peak and simultaneously obtain an
optimal definition of the off-pulse region and a stringent upper
limit on its amplitude of $<1.6\%$ of the mean flux.  The fine-grained
likelihood and BB partition together capture all of the information in
the pulse profiles in a simple visual representation.  

\section{Power Spectra}
\label{sec:ps}


A small but important class of LAT sources include $\gamma$-ray
binaries such as LS\,I $+$61$\arcdeg$ 303 and LS~5039.  These sources,
with an O or Be-type stellar companion, are known as microquasars and
are strong emitters of $\gamma$ rays modulated at the orbital period.
In some systems, the nature of the compact object is still unknown and
may either be a neutron star or a stellar-mass black hole.  In most
models the orbital modulation is governed by the dependence of Doppler
boosting on the viewing angle; see \citet{Dubus13} for an overview.

Because $\gamma$-ray binaries are exceedingly scarce, the discovery of
new systems plays an important role in understanding their nature, and so
motivates the development of more sensitive search techniques.
\citet{Corbet07} devised a method of ``exposure weighting'' to
measure power spectra in 
data sets with very inhomogeneous exposure, and later incorporated
photon weights \citep{Corbet10} to increase the sensitivity.  These
developments were key to the discovery of new $\gamma$-ray binaries such as 1FGL\,J1018.6$-$5856 \citep{Ackermann12_j1018} and CXOU
J053600.0$-$673507 in the LMC \citep{Corbet16}.  Here, we demonstrate
a connection between these formulations and the methods developed here.  The new formalism
puts
the search methods on a rigorous footing, offers an analytic
normalization for the control of trials factor, and in some cases improves
sensitivity (e.g. \S\ref{sec:reweighting}).

We begin by expanding the normalized flux densities for the source and background over time using a Fourier series:
\begin{align*}
\alpha(t) = &\sum_{l=1}^N \alpha_{cl} \cos(l\phi) + \alpha_{sl} \sin(l\phi)\\
\beta(t) = & \sum_{l=1}^N \beta_{cl} \cos(l\phi) + \beta_{sl} \sin(l\phi)
\end{align*}
with the sum extending up to the Nyquist frequency for the data and
$\phi\equiv2\pi t/T$ with $T$ the length of the data set.  In the
following derivation, we assume that the modes are independent and for
brevity concentrate on a single source and background mode, taken
without loss of generality to be 
$\alpha_{c1}$ and $\beta_{c1}$.  In most analyses, this assumption of
independence is a good one, but we assess the effect of the window
function further below.  Now, defining $(1-w_i)\equiv\bar{w}_i$, Eq.
\ref{eq:simple_like} becomes

\onecolumngrid
\begin{align*}
\log\mathcal{L} =& \sum_k \bigg[ \sum_{i\in k} \log
\big[1+w_{i}\big(\alpha_{c1}\cos(\phi_k)+\ldots\big) +
\bar{w}_i\big(\beta_{c1}\cos(\phi_k)+\ldots\big)\big]\bigg]\\
&- S_k[\alpha_{c1}\cos(\phi_k)+\ldots]-B_k[\beta_{c1}\cos(\phi_k)+\ldots]
\end{align*}
This expression can be further simplified if $\alpha_{c1} w_i\ll1$
and/or $\beta_{c1} \bar{w}_i\ll1$.  $\alpha$ describes the fractional
modulation associated with a mode, so can be small when either the
overall modulation relative to the source intensity is low, or when a
complex waveform requires small contributions from many modes.  As an
example, a typical low-order mode of a pulsar waveform might have
$\alpha\sim0.1$.  The
weights $w_i$ are small when the source is faint relative to the
background, which is the case for almost all \textit{Fermi} sources.
Thus, the product $\alpha_{c1}w_i$ is almost always $\ll1$.  For
background photons due to other sources, similar arguments apply.
For strong diffuse background sources, $\bar{w}_i\approx1$ is typical, but such
backgrounds are not strongly modulated, so $\beta_{c1}\ll1$
is an excellent approximation.  Residual particle background, on the
other hand, may be strongly modulated as \textit{Fermi} orbits the earth, but the fraction of such events that
survive background rejection is small such that $\bar{w}_i\ll1.$  Thus, we expand the logarithm and
suppress subscripts for other modes to find
\begin{align*}
\log\mathcal{L} &\approx \sum_k \sum_{i\in k}
\cos(\phi_k)\bigg[w_{i}\alpha + \bar{w}_i\beta-\alpha S_k-\beta
B_k\bigg] -\cos^2(\phi_k)\bigg[\frac{1}{2}w_{i}^2\alpha^2 +
w_i\bar{w}_i\alpha\beta+\frac{1}{2}\bar{w}_i^2\beta^2\bigg]\\
&\equiv \sum_k \cos(\phi_k)\bigg[W_k\alpha + \bar{W}_k\beta-\alpha
S_k-\beta B_k\bigg] -\cos^2(\phi_k)\bigg[\frac{1}{2}WW_k\alpha^2 +
W\bar{W}_k\alpha\beta+\frac{1}{2}\bar{W}\bar{W}_k\beta^2\bigg]\\
&\equiv \alpha \langle W-S\rangle_c + \beta \langle\bar{W} -B\rangle_c
-\frac{1}{2}\alpha^2\langle\langle WW \rangle\rangle_c
-\alpha\beta\langle\langle W\bar{W}\rangle\rangle_c
-\frac{1}{2}\beta^2\langle\langle \bar{W}\bar{W}\rangle\rangle_c,
\end{align*}
where we have grouped weights in the same cell into the series $W_k$
and $\bar{W}_k$ and denoted the cosine-weighted sums over cells with
e.g. $\sum_k \cos(\phi_k) W_k \equiv \langle W \rangle_c$ and 
$\sum_k \cos^2(\phi_k) W\bar{W}_k \equiv \langle\langle W\bar{W} \rangle
\rangle_c$.  (We suppress these subscripts until needed.)  Finally, if we differentiate the likelihood with respect to $\alpha$ and $\beta$ and solve the resulting system of equations, we find the maximum likelihood estimators
\begin{align}
\hat{\alpha}=\frac{\langle\langle\bar{W}\bar{W}\rangle\rangle\langle
W-S\rangle -\langle\langle W\bar{W}\rangle\rangle\langle \bar{W}-B\rangle}{\langle\langle WW\rangle\rangle\langle\langle \bar{W}\bar{W}\rangle\rangle-\langle\langle W\bar{W}\rangle\rangle^2} &&
\hat{\beta}=\frac{\langle\langle{W}{W}\rangle\rangle\langle \bar{W}-B\rangle -\langle\langle W\bar{W}\rangle\rangle\langle W-S\rangle}{\langle\langle WW\rangle\rangle\langle\langle \bar{W}\bar{W}\rangle\rangle-\langle\langle W\bar{W}\rangle\rangle^2}.
\end{align}
On the other hand, if we assume that $\beta=0$ (steady background) or $\alpha=0$ (steady source), we obtain the simple estimators
\begin{align}
\hat{\alpha_0}=\frac{\langle W-S\rangle}{\langle\langle WW\rangle\rangle} &&
\hat{\beta_0}=\frac{\langle \bar{W}-B\rangle}{\langle\langle \bar{W}\bar{W}\rangle\rangle}.
\end{align}
Finally, we can insert these estimators back into the second-order
likelihood to find the change in the log likelihood relative to the
no-modulation null case,
$\delta\log\mathcal{L}=\log\mathcal{L}(\alpha=\hat{\alpha},\beta=\hat{\beta})-\log\mathcal{L}(\alpha=\beta=0)$:
\begin{equation}
\label{eq:deltalogl_ps}
2\,\delta\log \mathcal{L}\approx P_{c}\equiv
\frac{\langle\langle\bar{W}\bar{W}\rangle\rangle\langle W-S\rangle^2
-2\langle\langle W\bar{W}\rangle\rangle\langle W-S\rangle\langle
\bar{W}-B\rangle + \langle\langle WW\rangle\rangle \langle \bar{W}-B\rangle^2}{\langle\langle WW\rangle\rangle\langle\langle \bar{W}\bar{W}\rangle\rangle-\langle\langle W\bar{W}\rangle\rangle^2}
\end{equation}
\twocolumngrid
\begin{figure*}
\centering
\includegraphics[angle=0,width=0.98\textwidth]{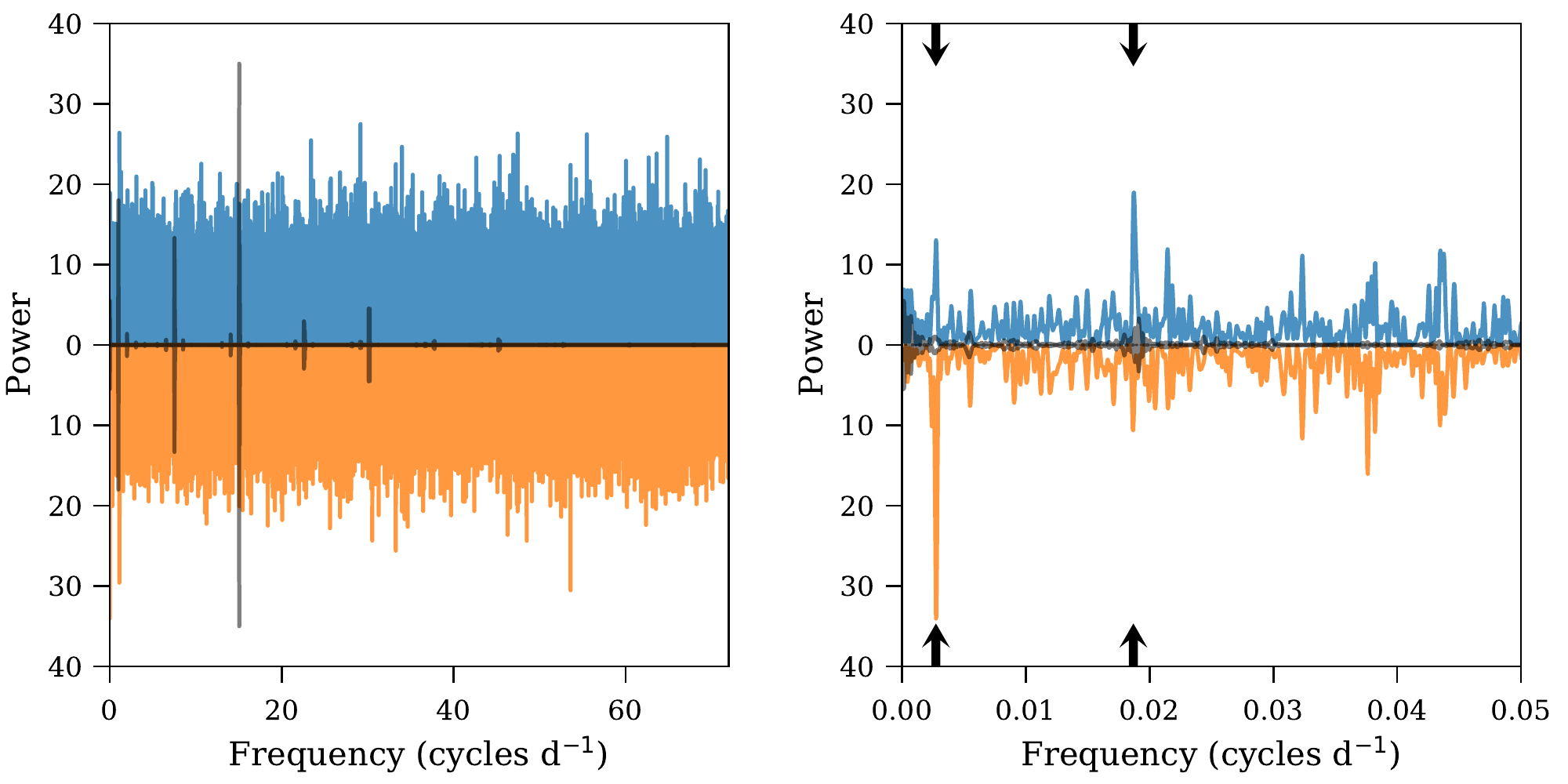}
\caption{\label{fig:geminga_ps}The power spectrum of Geminga.
(left) The full power spectrum is shown up to the limiting frequency
of 72 cycles\,d$^{-1}$, with the background-fixed estimator ($P_0$)
shown in orange below the x-axis and the profile likelihood ($P_1$) estimator in
blue, above.  The black trace gives the window function for the
exposure (see main text).
(right) A focus on low-frequency signals reveals the only two
systematic effects. Power in $P_0$ at a period of 1\,yr
(left black arrows) is likely caused by the sun, while power
in $P_1$ at the S/C precessional frequency (right arrows)
is probably induced by small inaccuracies in the exposure correction.}
\end{figure*}

\noindent We have defined $P_c$, the estimator for the power in the
``cosine'' mode, and there are analogous estimators and an expression
for $P_s$ with sine-weighted sums.  Combining these quantities gives
two estimators for the power spectrum, $P_0$ with the background
fixed to its time-averaged value ($\beta(t)=0$), and $P_1$ with
it fixed to its maximum likelihood value:
\begin{align}
\label{eq:ps0}
P_0 &= \frac{\langle W-S\rangle_c^2}{\langle\langle WW\rangle\rangle_c} + 
\frac{\langle W-S\rangle_s^2}{\langle\langle WW\rangle\rangle_s},\\
\label{eq:ps1}
P_1 &= P_c+P_s-\frac{\langle \bar{W}-B\rangle_c^2}{\langle\langle
\bar{W}\bar{W}\rangle\rangle_c}-\frac{\langle
\bar{W}-B\rangle_s^2}{\langle\langle \bar{W}\bar{W}\rangle\rangle_s}.
\end{align}
By Wilke's Theorem, both of these quantities are distributed as
$\chi^2_2$ in the null hypothesis.  We thus see that this estimator,
with a mean of 2, satisfies Leahy normalization \citep{Leahy83}.  The
numerators of $P_0$, $\langle W-S\rangle_c^2$ and $\langle
W-S\rangle_s^2$, are the exposure-weighted power spectrum estimators
of \citet{Corbet07}, so we see that the maximum likelihood estimator
$P_0$ is a generalization that is naturally Leahy normalized.
To this, $P_1$ adds the
ability to absorb background fluctuations.  Moreover, this
formulation suggests the evalulation of these sums using the Fast
Fourier Transform, and indeed, by using suitable identities, $P_0$ and
$P_1$ can be evaluated for all relevant frequencies by evaluating 2
and 5 FFTs, respectively.  For $P_0$, this formulation is
\begin{align}
\label{eq:ps_fft}
P_0(\nu)/2=&\frac{\mathrm{Re}[\mathcal{F}(W-S)(\nu)]^2}{\mathrm{Re}[\mathcal{F}(W^2)(2\nu)+\mathcal{F}(W^2)(0)]}\\\nonumber -& \frac{\mathrm{Im}[\mathcal{F}(W-S)(\nu)]^2}{\mathrm{Re}[\mathcal{F}(W^2)(2\nu)-\mathcal{F}(W^2)(0)]},
\end{align}
with $\mathcal{F}$ denoting a Fourier transform. The expression for $P_1$ is too lengthy to include here, and we
refer the reader to the implementation in \texttt{godot} for details.  These
formulations yield a substantial computational benefit, and as we show
in the examples below, enable construction of power spectra with
time scales as short as 2 minutes.  The evaluation time is perhaps
10\,s on a modest CPU.

We now give a series of examples demonstrating the manifestation of
various types of source signal in power spectra computed according to
Eqs. \ref{eq:ps0} and \ref{eq:ps1}.  As before we first test for
systematics by considering sources lacking \textit{bona fide}
astrophysical variability on these time scales.  Such sources should
be characterized by uniform power spectra, and the departure from such
a spectrum indicates a possible discovery, e.g. a line in the power
spectrum for a periodic source, a slightly wider feature for a
quasi-periodic oscillation, or colored noise for the stochastic
variability of an active galactic nucleus (AGN).  First, we consider
Geminga.  The pulsed emission (Fig.  \ref{fig:fig4}) is modulated at
the spin period of 237\,ms, far faster than the time scales we probe
here.  We binned the exposure into 300-s intervals, which allows
computation of frequencies of up to 72\,d$^{-1}$, or 20 minutes.
(Although the Nyquist frequency for 300-s sampling is 144\,d$^{-1}$,
the FFT formulation in Eq. \ref{eq:ps_fft} requires the evaluation of
FFTs up to twice the maximum frequency of interest, so only the lower
half of the Nyquist band is available.)  The power spectra, obtained
both with the fixed- ($P_0$) and free-background ($P_1$) assumptions,
appear in Figure \ref{fig:geminga_ps}.  There are about $10^6$
independent frequency samples in this spectrum, and it is almost
perfectly white,.  Moreover, the
samples follow the expected $\chi^2_2$ distribution, and the maximum
value of $P=27.8$ is expected to occur by chance in a sample this large with frequency
$\mathcal{O}(1)$.

This nearly featureless spectrum is remarkable considering that the
exposure varies periodically on a wide range of scales, as does the
background.  The ``window function'', the power spectrum of the
exposure towards Geminga, appears as a black trace superimposed on the
power spectrum, arbitrarily scaled such that the maximum power near
the S/C orbital frequency ($f_{sc}=15.1$\,d$^{-1}$, about
95.4\,minutes) is 35.  The window function has strong variation at the
S/C precessional frequency ($f_{prec}=0.019$\,d$^{-1}$, about 53.4\,d)
and at $f_{sc}$ and its harmonics.  The background from residual
particles, which contributes variability but does not influence the
window function, varies principally on 1-day time scales, as this is
roughly the time required to make a complete circuit of the
magnetosphere.  We refer the reader to the FSSC
documentation\footnote{\url{
https://fermi.gsfc.nasa.gov/ssc/data/analysis/LAT_caveats_temporal.html}}
for a more complete description of variability time scales.

The small signal in $P_0$ at 1\,yr$^{-1}$
($f=2.7\times10^{-3}$\,d$^{-1}$) is likely associated with the annual
passage of the sun, a modestly strong and extended $\gamma$-ray
source, by Geminga, at ecliptic latitude $-5.4^{\circ}$. Because the
sun's extended emission is largely degenerate with the steady
background, the contamination vanishes when using the profile
likelihood  ($P_1$) estimator for the power spectrum.  On the other
hand, the $P_1$ signal at $f_{prec}$ likely results
from very small inaccuracies in the exposure calculation.

\begin{figure}
\centering
\includegraphics[angle=0,width=1.00\linewidth]{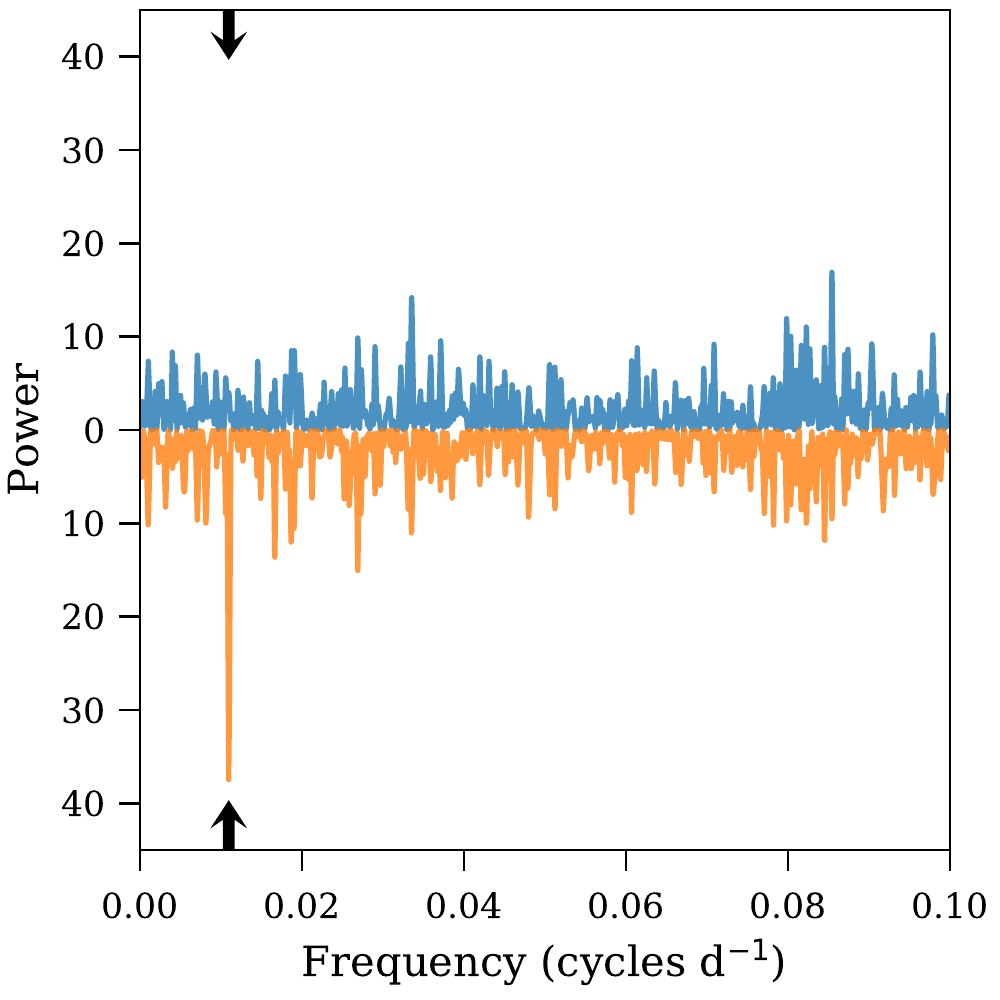}
\caption{\label{fig:j0823_ps}The power spectrum for
3FGL\,J0823$-$4205c, focused on low frequencies.  The lower, orange
trace, $P_0$, shows a strong peak at 0.011\,d$^{-1}$ (black arrows; 4 cycles
per year, see main text) while the upper blue trace, $P_1$, shows no
signal here or at high frequencies (not pictured).}
\end{figure}

Next, we consider 3FGL\,J0823.3$-$4205c, a much weaker source that
lies only 3$\arcdeg$ away from Vela (PSR~J0835$-$4510), the brightest
persistent GeV $\gamma$-ray source.  The power spectrum is featureless
at all frequencies save for the portion shown in Figure
\ref{fig:j0823_ps}, which reveals a signal in $P_0$ at 4\,yr$^{-1}$
($f=0.011$\,d$^{-1}$).  This artefact stems from the four-fold
symmetry of the \textit{Fermi}-LAT point-spread function.  Over the
course of a year, the typical azimutal angle at which Vela is viewed
completes a full rotation.  The concomitant rotation of the projected
PSF on the sky causes the background contribution from Vela to vary
annually, and the four-fold symmetry leads to the 4\,yr$^{-1}$ signal.
As with the solar contamination of Geminga, $P_1$ is insensitive to
this effect.

\begin{figure*}
\centering
\includegraphics[angle=0,width=0.98\linewidth]{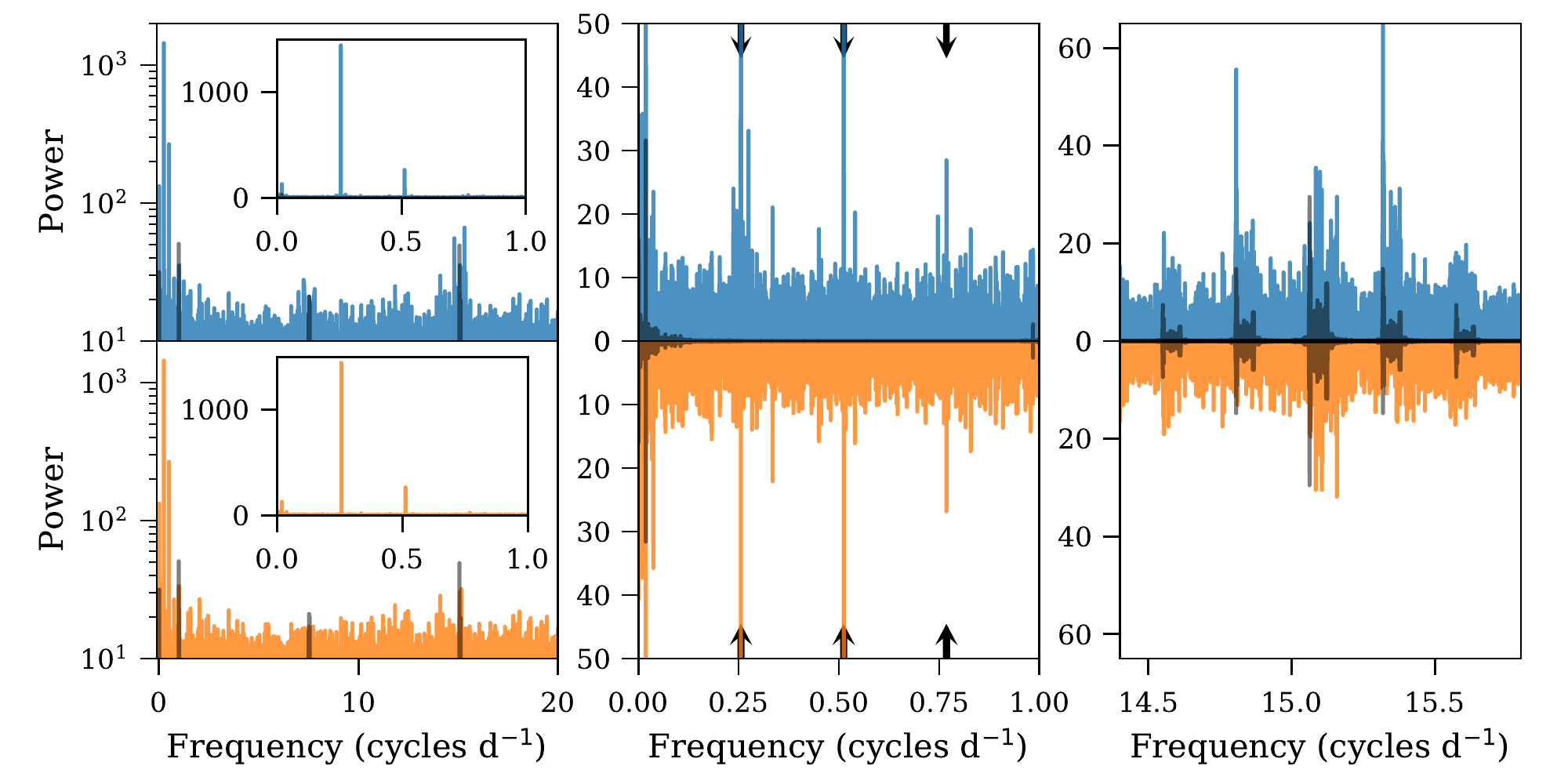}
\caption{\label{fig:ls5039_ps}The power spectrum of LS~5039.  The top
panels show the ``raw'' spectrum (blue), while the bottom panels
(orange) show
a version with time-domain signal subtraction, which reduces spectral
leakage.  (left)
The spectrum over a large section of bandwidth, with spectral leakage
visible at the S/C orbital frequency $f_{sc}=15.1$\,d$^{-1}$.  The inset shows the strength of
the LS~5039 modulation in the fundamental and first harmonic of the
orbital frequency $f_{ls}=0.26$\,d$^{-1}$.  (middle) A closer view of
spectral leakage around $f_{ls}$ and its harmonics (marked with arrows).  The leakage is essentially a copy of the window function produced by the \textit{Fermi} orbit.  (right)
Spectral leakage from $f_{ls}$ to $f_{sc}$.  The window function is shown in black and is
reproduced, scaled, and shifted by $f_{ls}$ and $2f_{ls}$,
respectively, to demonstrate the origin of the leakage.  The time
domain method removes almost all of the power associated with leakage
from the fundamental, while the broad peak at $f_{sc}$, due to low-frequency variation of LS~5039, remains, as does
power associated with $2f_{ls}$.}
\end{figure*}

These tests of steady sources are important for the \textit{discovery}
of modulations, i.e. excess power for an object with an
otherwise-white underlying spectrum.  We now consider characterization
of sources with known modulation and the effects of the window
function on its measurement.  Quite generally, power intrinsic to a
given frequency will be redistributed over the observed band;
specifically, this redistribution is given by the convolution of the
true power spectrum with the Fourier transform of the window function.
For \textit{Fermi}, with its discontinous viewing periods, this
manifests as a series of sinc$^2$-like peaks clustered around
$f_{prec}$, $f_{sc}$, and their harmonics (see Figures
\ref{fig:geminga_ps} and \ref{fig:ls5039_ps}). 

To illustrate the various features, we consider the power spectrum of
LS~5039, a strong $\gamma$-ray binary \citep[e.g.][]{Abdo09_ls5039}
with an orbital frequency $f_{ls}=0.26$\,d$^{-1}$ (period 3.9\,d).
The power spectrum appears in Figure \ref{fig:ls5039_ps} and is
dominated by the orbital modulation, clearly visible in the upper left
hand panel, with large power at $f_{ls}$ and $2\times f_{ls}$.  This
power ``leaks'' to other frequencies ``through'' the window function,
and this is clear from the main panel showing frequencies up to
20\,d$^{-1}$.  The upper blue trace is the raw power spectrum (Eq.
\ref{eq:ps0}), and there is clear power in excess of the white noise
floor near $f_{sc}$ and $f_{sc}/2$ ($\sim$15.1\,d$^{-1}$ and
$\sim$7.5\,d$^{-1}$), where the window function also has power.  The
middle panel highlights spectral leakage around $f_{ls}$ and its
harmonic; a single line has become a forest of power.  Finally, in the
upper right panel, we highlight the power that has appeared near
$f_{sc}$.  True low-frequency power associated with secular variation
of LS~5039 is convolved with the window function, producing a broad
peak at $f_{sc}$.  On the other hand, the two LS~5039 peaks, being
essentially delta functions, produce an exact copy of the window
function at $f_{sc}\pm f_{ls}$ and $f_{sc}\pm2f_{ls}$.  To illustrate
this, we have reproduced the window function and shifted it to these
frequencies.

To reduce this spectral leakage, it is possible to remove the strong
orbital modulations in the time domain \textit{before} evaluating Eq.
\ref{eq:ps0}, and then restoring the power afterwards.  We do this by
directly fitting a sinusoid at $f_{ls}$ using Eq. \ref{eq:simple_like}
with the 300s segments and updating the predicted counts
$S_k$ with the modulation before computing $P_0$ and $P_1$.  These results
appear in the lower orange traces.  Generally, the power spectrum is
unchanged in regions where the window function is minimal (exposure
variations washed out), but substantially improved near problematic
frequencies.  This approach is critical in the analysis of strongly
pulsed sources; see, e.g., the analysis of an ultraluminous X-ray
pulsar \citep{Wilson-Hodge18}.


\begin{figure}
\centering
\includegraphics[angle=0,width=0.98\linewidth]{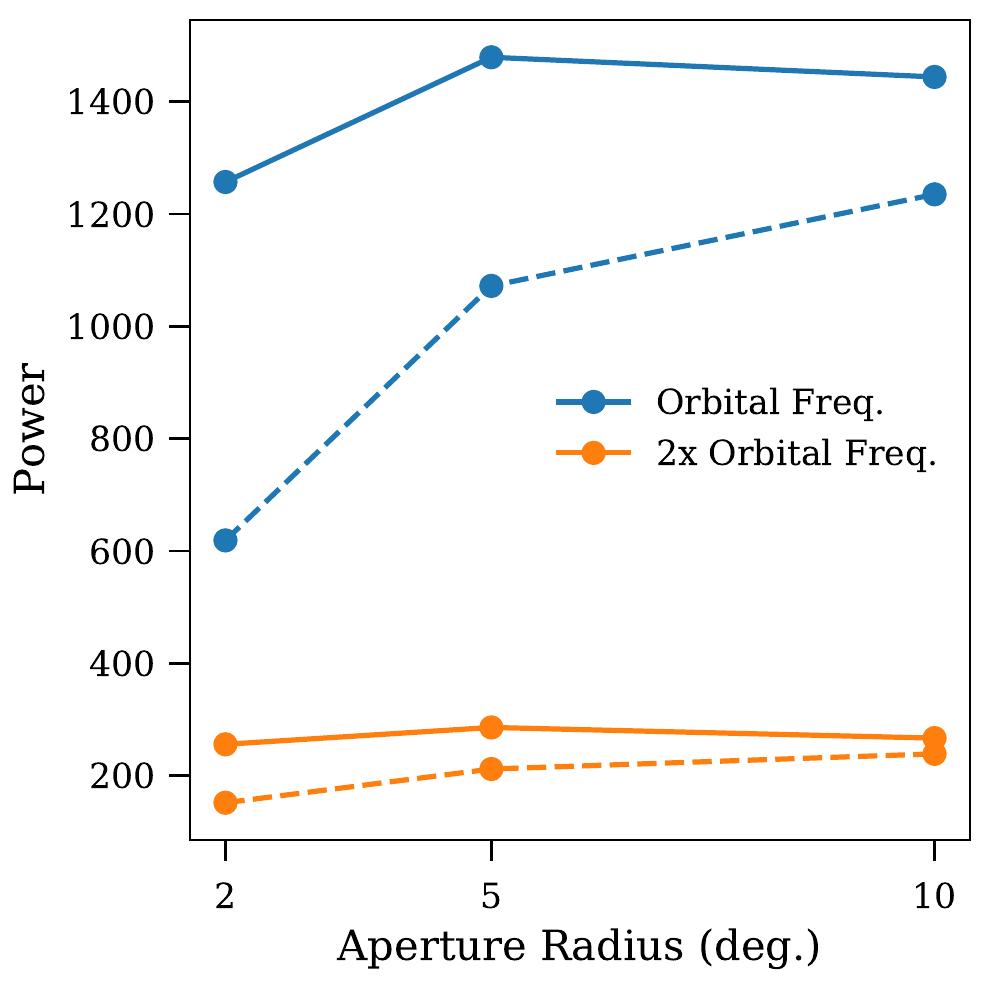}
\caption{\label{fig:ls5039_aperture}The power obtained at the fundamental
(blue) and first harmonic (orange) orbital frequency for LS~5039 when
using $P_0$/Eq. \ref{eq:ps0} (solid) and $P_1$/Eq. \ref{eq:ps1} (dashed).}
\end{figure}

Because the orbital modulation provides a clear way to distinguish the
source from the background, LS~5039 provides a nice way to study the
dependence of our two power spectrum estimators $P_0$ (Eq.\,\ref{eq:ps0},
with constant background), and $P_1$ (Eq. \ref{eq:ps1}, with variable
background) on ROI size.  Generally we
expect two effects: (1) a larger ROI should contain more source
photons than a smaller one, until the ROI fully encloses the PSF;
(2); the background will be degenerate with the source intensity until
the ROI is large enough to ``resolve'' the source.
\citet{Kerr11} showed that for most \textit{Fermi} pulsars a
2$\arcdeg$ ROI was sufficient to garner most of the signal, as
even though the LAT PSF is larger than this at low energies, the
signal-to-background ratio drops rapidly outside of the PSF core.
In Figure \ref{fig:ls5039_aperture} we show $P_0$ and $P_1$ at
$f_{ls}$ and $2\times f_{ls}$ (see Figure \ref{fig:ls5039_ps}, left panel
inset) as a function of ROI size, with two traces comparing the
two estimators.  Thus, we indeed see that most of the signal strength is
already present in a 2$\arcdeg$ ROI, but that jointly fitting the
background reduces that signal substantially, by about 50\%
for a small ROI.
On the other hand, for a 10$\arcdeg$ ROI, the difference is
modest.  Thus, the analyzer can choose between sensitivity, resilience
to background variation, and computational burden in choosing the
best-suited ROI.


\subsection{Barycentric Corrections} For binaries with periods shorter
than about an hour, the smearing (up to 500\,s) of the signal due to
light travel time from the barycenter to the observatory can dilute
the significance of the binary signal when folded over long time
spans.  To mitigate this, we provide an option in the software to
produce the binned weights and exposure time series in Barycentric
Dynamical Time (TDB) at the barycenter.  First, we convert a coarse
(about 10-minute resolution) time series of topocentric times to
barycentric times and then use these points as knots to interpolate
arbitrary topocentric times to barycentric times.  We then
establish a uniform series of bin edges in the barycentric times and
map these times back to the topocenter, which allows us to assign
photons/weights to the correct bins and to compute the exposure within
each uniform barycentric bin.  We refer the reader to the
\texttt{godot} software for implementation details.

\begin{figure*}
\centering
\includegraphics[angle=0,width=0.98\linewidth]{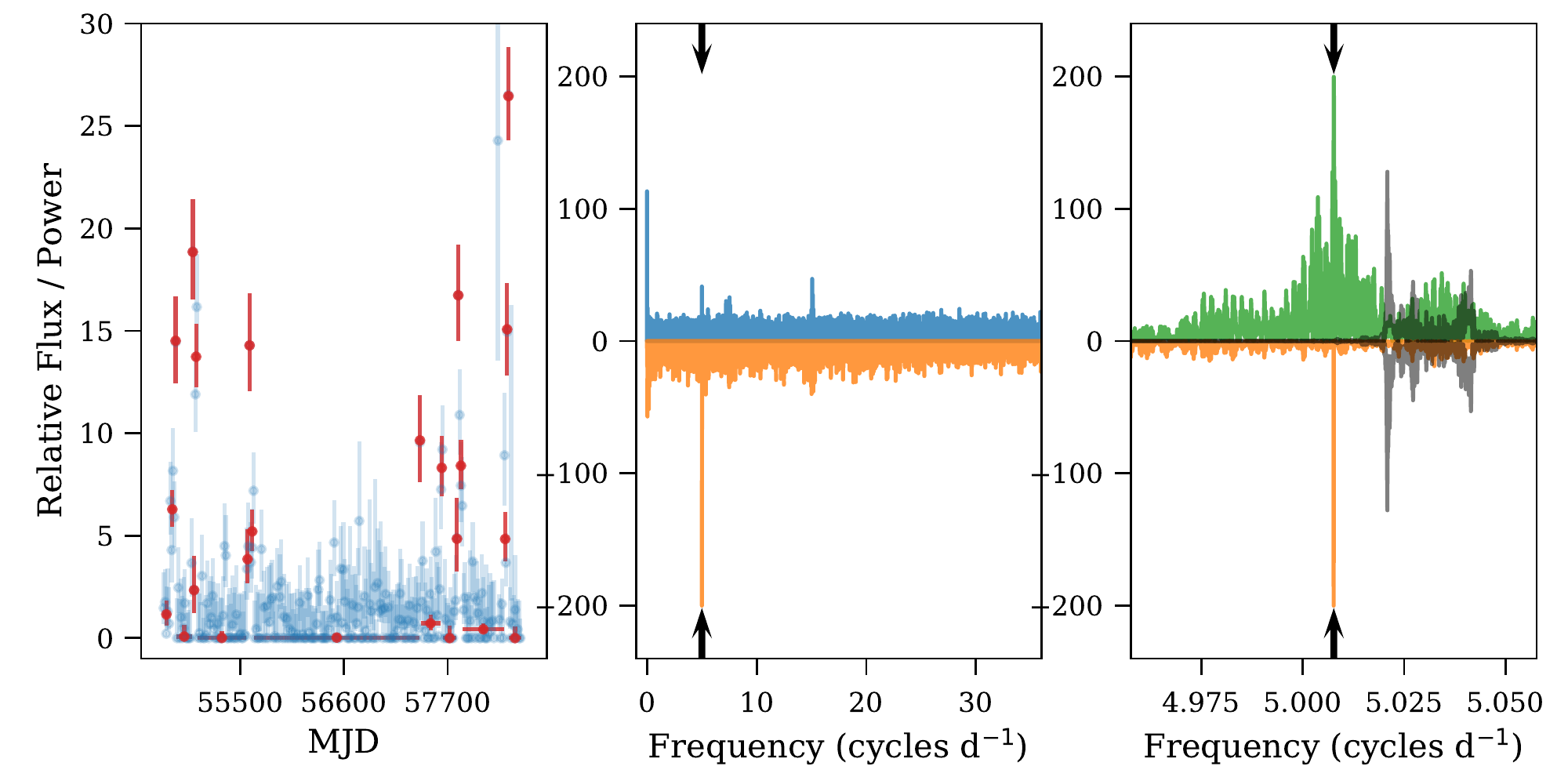}
\caption{\label{fig:cygx3}A modulation analysis of Cygnus X-3.  (left)
The 14-d light curve (blue, light points) and Bayesian blocks
estimators (red, heavier points)  for a
piecewise-constant light curve.  These light curves are used for
photon reweighting.  (middle) The power spectrum without (top, blue)
and with (bottom, orange) reweighting.  The arrows indicate
the Cygnus X-3 orbital frequency.  (right) A narrow band around the
$f_{cyg}$, showing the re-weighted power spectrum
without (top, green) and with (bottom, orange) spectral leakage reduction.  
The gray trace
shows the Fermi window function shifted to 1/3 of its original
frequency.}
\end{figure*}

\section{Reweighting}
\label{sec:reweighting}

Finally, we consider the ``reweighting'' of photons.  Such reweighting
is appropriate when, e.g., one has obtained weights from a long period
of time but wishes to analyze a short interval where the flux of the
source of interest is substantially different to the mean, e.g. a
flaring blazar.  In this case, if $\alpha$ is the ratio of the source
fluxes, then a new set of weights $w'$ can be obtained from the
original weights $w=s/(s+b)$:
\begin{equation}
\label{eq:reweight}
w' = \frac{\alpha\,s}{\alpha\,s + b}=
\frac{\alpha\frac{s}{s+b}}{\alpha\frac{s}{s+b} + \frac{b}{s+b}} = \frac{\alpha\,w}{\alpha\,w + (1-w)},
\end{equation}
and likewise the total expected source counts should
be scaled by $\alpha$.

The adoption of reweighting solves a lingering problem in the
application of photon weights to the search for binaries, namely that
the use of photon weights worked poorly for sources with long-term
variability.  Briefly, the exposure-weighting technique of
\citet{Corbet07} uses the mean photon rate to scale the exposure
correctly, so it adapts naturally to higher source rates.  On the
other hand, the use of a fixed set of weights effectively deweighted
source photons during high-flux states.  Here, we show how reweighting
naturally accounts for slow source variability and gives optimal
sensitivity for modulation searches.

For our example we take the microquasar Cygnus X-3.  The detection of
$\gamma$-ray modulation at its orbital frequency
\citep[$f_{cyg}=5.0076692$\,d$^{-1}$ at
MJD 56561,][]{Bhargava17} during radio and $\gamma$-ray flares was an
important early result from the LAT \citep{Abdo09_cygx3} and a
particular technical challenge given that
$f_{cyg}$$\approx$$f_{sc}/3$, raising questions of systematic error.
Distinguishing the two frequencies requires a long data span to
provide good resolution in the power spectrum; on the other hand,
Cygnus X-3 flares last typically only a few weeks.  In previous LAT
catalogs, Cygnus X-3 was too faint (and too soft) to be detected in
time-averaged analysis, and it is only with the long data span and new
capabilities of the Pass 8 event reconstruction that it is firmly
detected in 8-year source lists.

To obtain a long, optimal power spectrum we have combined the methods
of \S\ref{sec:lc} and \S\ref{sec:ps}.  First, we divided the data into
14-day cells and ran the BB algorithm to obtain a piecewise constant
estimator for $\alpha(t)$ on long timescales.  This time series
appears in the left panel of Figure \ref{fig:cygx3}.  We subsequently
applied Eq. \ref{eq:reweight} using the appropriate value of $\alpha$
within each partition, taking care to also scale the total expected
source counts (``S'').  Finally, we compute the power spectrum using
Eq. \ref{eq:ps0}.  In the middle panel, we show the power spectrum
without reweighting (using the time-averaged weights) with the upper
blue trace and that with reweighting (and spectral leakage reduction)
with the lower orange trace.  The reweighting increases the S/N by a
factor of about 5, and also reduces artefacts by absorbing the
low-frequency variation into the scale factor.  In the rightmost
panel, we show a narrow band around $f_{cyg}$.
Here, the top trace shows the reweighted signal without spectral
leakage reduction, while the bottom orange trace results from removing
the signal at $f_{cyg}$ in the time domain and restoring it.  The
reweighting decreases the uniformity of the window function and
broadens the range over which spectral power is leaked, making 
time-domain subtraction particularly effective.  Finally, we show the
window function shifted to from $f_{sc}$ to $f_{sc}/3$ to demonstrate
how close $f_{cyg}$ is to $f_{sc}/3$ but also that they are clearly
resolved.  This is the strongest (and cleanest) $\gamma$-ray detection
of the orbital modulation of Cygnus X-3 to date.



\section{Discussion and Summary}
\label{sec:discussion}

We have introduced a method of \textit{retrospective} analysis that
approximates the full likelihood of \textit{Fermi}-LAT data.  Because
of its speed and simplicity, it is suitable for a wide range of
applications.  In particular, monitoring large numbers of sources
(e.g. AGN) can be done cheaply and with more sensitivity than simple
aperture photometry.  The methods are also well suited for probing the
shortest variability timescales, and we demonstrated that an
``omnibus'' test for variability can be achieved by using
short-in-time data cells along with the Bayesian blocks algorithm.  We
conclude the paper by considering a few additional applications.

To study faint pulsar wind nebula or supernova remnants, it is often
useful to ``gate'' out emission from bright pulsars by selecting data
only from the off-pulse \citep[e.g.][]{Grondin13,Li18}.  Rather than hard
cuts, analysts can apply Eq. \ref{eq:reweight} to time-averaged weights
derived for the pulsar using for $\alpha(t)$ the pulse profile.  Thus,
photons from the pulse peaks have a weight very close to 1, while
those from the off-pulse will be given a weight close to 0.  By taking
the inverse of these weights, a data set retaining as much information
as possible, but with the pulsar signal optimally diminished, is
obtained.

In addition to the longer-period $\gamma$-ray binaries discussed in
\S\ref{sec:ps}, a second class of sources emitting orbitally-modulated
$\gamma$-rays are eclipsing millisecond pulsars in tight binaries,
including the so-called black widows and red backs.  In these systems, the
pulsar wind can interact with the wind of either a compact, main
sequence, or evolved companion, with the shocked interface emitting
X-rays \citep[e.g.][]{Gentile14} and $\gamma$-rays
\citep[e.g.][]{An17}.  As with the massive star binaries, changing the
viewing angle with respect to the bulk flow of the wind produces a
modulation in spectrum and intensity.  Additionally, dense material
ablated from the stellar surface can eclipse radio emission from the
pulsar, and for nearly edge-on systems a direct eclipse of $\gamma$
rays can occur.  Although rare, detection of such an eclipse would
immediately identify the orbital period and constrain the system's
inclination.  \citet{Romani12} performed such a search using a
photon-weighted based likelihood and a template of a complete eclipse
to identify very marginal evidence for an eclipse of PSR~J2215$+$5135.
The methods here are sufficiently fast to enable a search of all \textit{Fermi} LAT
sources for such eclipses.


In this paper, we concentrated on characterizing variability or in
discovering variability in \textit{known} sources.  However, a clear
extension is the search for new variable sources, specifically those
that are too faint to detect in time-averaged analyses but are
strongly detected via flares.  (Cygnus X-3 is a good example, as are
many AGN.)  In order to find new sources, it suffices simply to
introduce a source over a series of test positions and compute a set
of weights for each one.  Provided the source is assigned a flux near
threshold (so as not to perturb the background weights), the methods
outlined here will work well, and e.g. the detection of variability
with Bayesian blocks can motivate a dedicated followup with a full
likelihood analysis.

Finally, we consider the applicability of the methods presented here
to other wavebands and instruments.  Although photon counting
instruments are available from the THz through UV, and such data
perfectably amenable to photon-weight analysis, the event rates are
too high to be practical.  On the other hand, imaging X-ray
instruments, particularly those that have traded angular resolution
for effective area, e.g. \textit{XMM-Newton}, \textit{NICER}, and the
future \textit{Athena} mission, are well suited.  Indeed, the large
collecting area and very good spectral resolution of \textit{Athena}
make fitting spectro-variability models with photon weights a very
promising application.  At still higher energies, coded aperture
telescopes used for hard X-ray observations have intrinsically high
backgrounds, and could benefit from weighted analyses.  Finally, in
order to characterize TeV sources, Cherenkov imaging telescopes must
separate particle showers produced by $\gamma$ rays and cosmic rays.
Residual cosmic ray events form a natural background and weights can
be incorporated in two ways: (1) with a probabilistic classification
based on properties of the shower (2) with astrophysical weighting
based on the distribution of position and energies of events in an
ROI, as done here.

In summary, we believe the methods introduced here hold great promise
for more sophisticated LAT analyses and for application to a number of
high-energy instruments both in operation and yet to come.

\acknowledgements
The \textit{Fermi} LAT Collaboration acknowledges generous ongoing
support from a number of agencies and institutes that have supported
both the development and the operation of the LAT as well as
scientific data analysis.  These include the National Aeronautics and
Space Administration and the Department of Energy in the United
States, the Commissariat \`a l'Energie Atomique and the Centre
National de la Recherche Scientifique / Institut National de Physique
Nucl\'eaire et de Physique des Particules in France, the Agenzia
Spaziale Italiana and the Istituto Nazionale di Fisica Nucleare in
Italy, the Ministry of Education, Culture, Sports, Science and
Technology (MEXT), High Energy Accelerator Research Organization (KEK)
and Japan Aerospace Exploration Agency (JAXA) in Japan, and the
K.~A.~Wallenberg Foundation, the Swedish Research Council and the
Swedish National Space Board in Sweden.
 
Additional support for science analysis during the operations phase
is gratefully acknowledged from the Istituto Nazionale di Astrofisica
in Italy and the Centre National d'\'Etudes Spatiales in France. This
work performed in part under DOE Contract DE-AC02-76SF00515.

Work at NRL is supported by NASA, in part by Fermi Guest Investigator
grant NNG19OB19A.  The author is grateful to Lucas Guillemot, Robin
Corbet, Benoit Lott, and Philippe Bruel for helpful discussion and
suggestions, to Toby Burnett for assistance in preparing data sets,
and to the anonymous referee for a detailed and constructive review
that substantially improved this work.

\facilities{Fermi}

\bibliographystyle{aasjournal}


\end{document}